\documentclass[prl,aps,twocolumn,preprintnumbers, showpacs, nofootinbib,superscriptaddress,notitlepage]{revtex4-1}
\usepackage{hyperref}
\usepackage{graphicx}
\usepackage{subfigure}
\usepackage{amsmath}
\usepackage{amssymb}
\usepackage{slashed}
\usepackage{amsfonts}
\usepackage{xcolor}
\usepackage{multirow}
\usepackage{array}
\usepackage{mwe}
\usepackage{mathrsfs}
\usepackage{comment}
\usepackage[normalem]{ulem}

\begin{document}
\title{Intrinsic Soft Function in the Continuum and Physical Mass Limit from Lattice QCD}

\collaboration{\bf{Lattice Parton Collaboration ($\rm {\bf LPC}$)}}

\author{\includegraphics[scale=0.10]{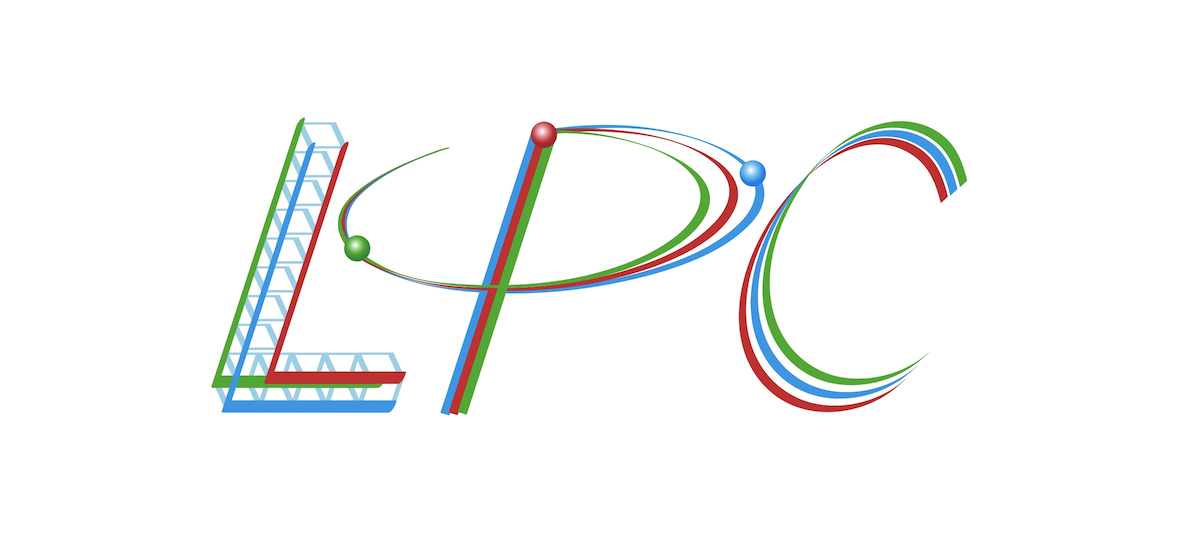}\\Zhi-Chao Gong}
\affiliation{Tsung-Dao Lee Institute, Shanghai Jiao Tong University, Shanghai 201210, China}
\affiliation{State Key Laboratory of Dark Matter Physics, Key Laboratory for Particle Astrophysics and Cosmology (MOE), Shanghai Key Laboratory for Particle Physics and Cosmology, School of Physics and Astronomy, Shanghai Jiao Tong University, Shanghai 200240, China}

\author{Min-Huan Chu}
\email{Corresponding author: minhuan.chu@amu.edu.pl}
\affiliation{Faculty of Physics and Astronomy, Adam Mickiewicz University, ul.\ Uniwersytetu Pozna\'nskiego 2, 61-614 Pozna\'{n}, Poland}

\author{Jun Hua}
\affiliation{State Key Laboratory of Nuclear Physics and Technology, Institute of Quantum Matter, South China Normal University, Guangzhou 510006, China}
\affiliation{Guangdong Basic Research Center of Excellence for Structure and Fundamental Interactions of Matter, Guangdong Provincial Key Laboratory of Nuclear Science, Guangzhou 510006, China}

\author{Xiangdong Ji}
\affiliation{Tsung-Dao Lee Institute and School of Physics and Astronomy, Shanghai Jiao Tong Unviersity, Shanghai 200240, China}

\author{Andreas Sch\"afer}
\affiliation{Institut f\"ur Theoretische Physik, Universit\"at Regensburg, D-93040 Regensburg, Germany}
\affiliation{Department of Physics, National Taiwan University, Taipei, Taiwan 106, China}

\author{Jin-Xin Tan}
\affiliation{Tsung-Dao Lee Institute, Shanghai Jiao Tong University, Shanghai 201210, China}
\affiliation{State Key Laboratory of Dark Matter Physics, Key Laboratory for Particle Astrophysics and Cosmology (MOE), Shanghai Key Laboratory for Particle Physics and Cosmology, School of Physics and Astronomy, Shanghai Jiao Tong University, Shanghai 200240, China}

\author{Han-Zhang Wang}
\affiliation{State Key Laboratory of Dark Matter Physics, Key Laboratory for Particle Astrophysics and Cosmology (MOE), Shanghai Key Laboratory for Particle Physics and Cosmology, School of Physics and Astronomy, Shanghai Jiao Tong University, Shanghai 200240, China}
\affiliation{Zhiyuan College, Shanghai Jiao Tong University, Shanghai 200240, China}

\author{Wei Wang}
\email{Corresponding author: wei.wang@sjtu.edu.cn}
\affiliation{State Key Laboratory of Dark Matter Physics, Key Laboratory for Particle Astrophysics and Cosmology (MOE), Shanghai Key Laboratory for Particle Physics and Cosmology, School of Physics and Astronomy, Shanghai Jiao Tong University, Shanghai 200240, China}
\affiliation{Southern Center for Nuclear-Science Theory (SCNT), Institute of Modern Physics, Chinese Academy of Sciences, Huizhou 516000, Guangdong Province, China}

\author{Jian-Hui Zhang}
\affiliation{School of Science and Engineering, The Chinese University of Hong Kong, Shenzhen, Guangdong, 518172, P.R. China}

\author{Mu-Hua Zhang}
\affiliation{Tsung-Dao Lee Institute, Shanghai Jiao Tong University, Shanghai 201210, China}
\affiliation{State Key Laboratory of Dark Matter Physics, Key Laboratory for Particle Astrophysics and Cosmology (MOE), Shanghai Key Laboratory for Particle Physics and Cosmology, School of Physics and Astronomy, Shanghai Jiao Tong University, Shanghai 200240, China}

\author{Qi-An Zhang}
\email{Corresponding author: zhangqa@buaa.edu.cn}
\affiliation{School of Physics, Beihang University, Beijing 102206, China}

\begin{abstract}
The intrinsic soft function is a universal nonperturbative ingredient of transverse-momentum-dependent (TMD) factorization. Despite substantial progress in lattice QCD studies, a determination with simultaneous control of the continuum, physical-pion-mass, and infinite-momentum limits has remained unavailable. In this work we present the first lattice determination incorporating all these limits within large-momentum effective theory, extracting the intrinsic soft function from a light-meson form factor and quasi-TMD wave functions. A key advance is the use of power-counting-guided axial-vector interpolators, which substantially improve the signal and enable a direct calculation at the physical pion mass. Combined with results from six ensembles spanning four lattice spacings, this improvement allows a precise determination extending to $b_\perp\simeq1.1~\mathrm{fm}$, while remaining consistent with perturbative predictions at short distances. The resulting transverse-distance dependence exhibits an approximately exponential decay, with the decay rate being characterized by an effective scale $\Lambda\simeq0.7~\mathrm{GeV}$ at $\mu=2~\mathrm{GeV}$, providing a quantitative benchmark for nonperturbative Wilson-line dynamics and essential input for precision TMD studies.
\end{abstract}

\maketitle

\textit{Introduction.}--
The soft function is a universal object in quantum chromodynamics (QCD), independent of both the underlying hard-scattering process and the external hadron states. As the vacuum matrix element of Wilson lines along two conjugate light‑like directions, it encodes the soft‑gluon infrared structure of the theory and mediates the real‑virtual cancellation that enables infrared‑finite, well‑defined measurements~\cite{Collins:1981va,Collins:1984kg,Ji:2004wu,Ji:2004xq}. It thus captures an essential component of long‑distance QCD dynamics. Within TMD factorization, the scale evolution of the hard, collinear, and soft contributions resums the large logarithms in the transverse-momentum spectrum, making the high-order determination of the soft function essential for precision phenomenology~\cite{Li:2016ctv,Wang:2018vgu}. Its rapidity dependence, governed by the Collins–Soper kernel, can be isolated by leaving behind the rapidity‑independent intrinsic soft function \(S_I(b_\perp,\mu)\)~\cite{Collins:1981va,Ji:2019sxk,LatticeParton:2020uhz}.

At short distances, the soft function is perturbatively calculable~\cite{Echevarria:2011epo,Chiu:2012ir,Echevarria:2015byo,Lubbert:2016rku,Li:2016ctv,Vladimirov:2016dll,Vladimirov:2017ksc,Moult:2018jzp}, with its small-$b_\perp$ behavior governed by renormalization and rapidity evolution. The intrinsic soft function probes the nonperturbative color dynamics of QCD through transverse correlations encoded in its Wilson-line structure. At hadronic transverse separations, these correlations provide a window into the long-distance dynamics underlying confinement. At large $b_\perp$, the intrinsic soft function is also related to the Collins--Soper kernel through the rapidity structure of the soft function~\cite{Ji:2019sxk}. This is precisely the region where fixed-order perturbation theory and the small-$b_\perp$ operator product expansion cease to be reliable, and a first-principles determination of the intrinsic soft function therefore requires lattice QCD. While the nonperturbative Collins--Soper kernel has been determined on the lattice with steadily improving systematic control~\cite{Shanahan:2020zxr,LatticeParton:2020uhz,Shanahan:2021tst,Schlemmer:2021aij,LatticePartonLPC:2022eev,LatticePartonCollaborationLPC:2022myp,LatticePartonLPC:2023pdv,Avkhadiev:2023poz,LatticeParton:2024mxp,LatticeParton:2025eui,Bollweg:2025ecn,Avkhadiev:2024mgd,Bollweg:2024zet,Liu:2024sqj,Avkhadiev:2025wps,Bollweg:2025iol,Tan:2025ofx,Francis:2026czb,Avkhadiev:2026xyf}, the intrinsic soft function, another essential nonperturbative ingredient of TMD factorization, remains considerably less constrained.

Because the light-like Wilson lines defining $S_I$ cannot be implemented directly in Euclidean lattice QCD, Large-momentum effective theory (LaMET) provides a framework to extract it from a large-momentum light-meson form factor~\cite{Ji:2019sxk}. In this approach, the form factor of two local currents separated by $b_\perp$ factorizes at large $P^z$ into the intrinsic soft function, a perturbative hard kernel, and two quasi-TMD wave functions, which can be calculated independently~\cite{Ji:2019sxk,Deng:2022gzi}. Following the exploratory tree-level calculation~\cite{LatticeParton:2020uhz}, this approach has been further developed through improved Dirac structures, one-loop matching, and studies of large-$b_\perp$ and finite-volume effects~\cite{Li:2021wvl,LatticePartonLPC:2023pdv,Alexandrou:2025xci,Zhao:2023ptv,Bollweg:2025iol}. Nevertheless, a simultaneous control of the continuum, physical-pion-mass, and infinite-momentum limits has remained elusive. In particular, our previous calculation was limited by the rapidly deteriorating signal-to-noise ratio of the form factor at large $P^z$ and $b_\perp$ with conventional pseudoscalar interpolators, making a precise determination in this kinematic regime challenging, especially at the physical pion mass. Although $S_I$ is a vacuum matrix element of Wilson lines, it retains a dependence on the sea-quark masses through the gauge-field configurations. The mild external-pion-mass dependence observed in previous calculations therefore does not establish the sea-quark-mass dependence of $S_I$, which requires a direct calculation with physical-mass ensembles.

To address this limitation, we exploit the distinct large-momentum scaling of the signal and noise in the light-meson form factor. We employ leading-power axial-vector interpolators whose overlap with the boosted meson grows linearly with $P^z$, while the noise receives no corresponding enhancement, consistent with the general spirit of LaMET power counting~\cite{Ji:2024oka}. The resulting improvement in precision at large momentum and transverse separation enables the first direct determination of the intrinsic soft function at the physical pion mass. Using six ensembles spanning four lattice spacings and pion masses down to the physical point, with three boosted momenta per ensemble, we simultaneously control the continuum, physical-pion-mass, and infinite-momentum limits, resolving $S_I$ up to $b_\perp\simeq1.1$~fm, deep in the nonperturbative region. The intrinsic soft function exhibits an approximately exponential decay in $b_\perp$, providing an effective nonperturbative transverse scale for investigating long-distance Wilson-line dynamics.


\textit{Theoretical framework.}--
The TMD soft function is defined as a vacuum matrix element of Wilson lines extending along two conjugate near-lightlike directions. In the large-rapidity limit, its rapidity dependence is governed by the Collins--Soper kernel, while the remaining rapidity-independent component defines the intrinsic soft function $S_I(b_\perp,\mu)$. The latter can be extracted through the large-momentum factorization of a specially constructed form factor of a light pseudoscalar meson~\cite{Ji:2019sxk}. For initial and final mesons carrying equal and opposite spatial momenta, $\vec{P}_1=-\vec{P}_2=(0,0,P^z)$, the normalized bare form factor is defined as
\begin{align}
F^{(0)}_{\Gamma}(b_\perp,P^z)
=\frac{\langle \pi(\vec{P}_2)|[\bar{q}_2\Gamma q_1](b_\perp)[\bar{q}_1\Gamma q_2](0)|\pi(\vec{P}_1)\rangle}{f_\pi^2\,\vec{P}_1\cdot \vec{P}_2},
\label{eq:def_form_factor}
\end{align}
where the two local currents are separated by a purely transverse distance $b_\perp$, with the flavor assignments selecting the connected contraction. Through Fierz rearrangement, both $F_{\gamma_\perp}+F_{\gamma_\perp\gamma_5}$ and $F_I-F_{\gamma_5}$ project onto the same leading-twist pion amplitude~\cite{Li:2021wvl,Deng:2022gzi}. However, the scalar and pseudoscalar currents require additional ultraviolet renormalization and have nonzero anomalous dimensions, resulting in a pronounced residual $P^z$ dependence at finite perturbative order~\cite{LatticePartonLPC:2023pdv}. We therefore employ the former combination, which provides a numerically cleaner suppression of higher-twist contamination~\cite{LatticePartonLPC:2023pdv}, and has been demonstrated numerically in the supplemental material~\cite{supplemental}.

In the large-momentum limit, the form factor follows the LaMET factorization \cite{Ji:2019sxk,Deng:2022gzi,LatticePartonLPC:2023pdv}
\begin{align}
    F_{\Gamma}(b_\perp,P^z,\mu)=S_I(b_\perp,\mu)\int dx_1 dx_2 H_\Gamma(x_1,x_2,P^z,\mu)\nonumber\\
    \times\widetilde{\Psi}^*(x_1,b_\perp,-P^z)\widetilde{\Psi}(x_2,b_\perp,P^z)+\mathcal{O}\left(\frac{\Lambda_{\rm QCD}^2, m_{\pi}^2}{(P^z)^2}\right),
    \label{eq:FF_fac}
\end{align}
where $H_{\Gamma}$ is the perturbative hard kernel encoding the short-distance scattering and $\widetilde{\Psi}$ is the renormalized quasi-TMD wave function describing the collinear dynamics of the incoming and outgoing mesons~\cite{Ji:2021znw,LatticeParton:2023xdl,Liu:2019urm,Liu:2018tox,LatticeParton:2024vck,Deng:2023csv}. The quasi-TMD wave functions are obtained with the same operator construction, renormalization, and analysis strategy as in the recent calculation~\cite{Tan:2025ofx}, which employed five of the six ensembles used here and implemented a state-of-the-art hybrid renormalization scheme. This scheme combines the nonperturbative subtraction of Wilson-line power divergences with a self-renormalization treatment of the remaining logarithmic divergence and conversion to the $\overline{\rm MS}$ scheme, providing improved control of ultraviolet systematics that was absent in previous lattice determinations of the intrinsic soft function. We refer to Ref.~\cite{Tan:2025ofx} for details of the operator definition, renormalization, and large-momentum analysis.

To extract the form factor, we first construct pion two-point correlation functions using pseudoscalar ($\gamma_5$) or axial-vector ($\gamma_\mu\gamma_5$) interpolators to determine the pion energies and overlap factors. We then construct three-point correlation functions with two local currents, whose Dirac structures include $\gamma_\perp$ and $\gamma_\perp\gamma_5$, separated by $b_\perp$. Dividing the three-point correlation function by the corresponding two-point functions removes the external-state overlap factors and the leading Euclidean time dependence. Fitting the ratios over multiple source-sink separations then yields the pion matrix element and the corresponding form factor.

\begin{figure}[tb]
\centering
\includegraphics[width=\columnwidth]{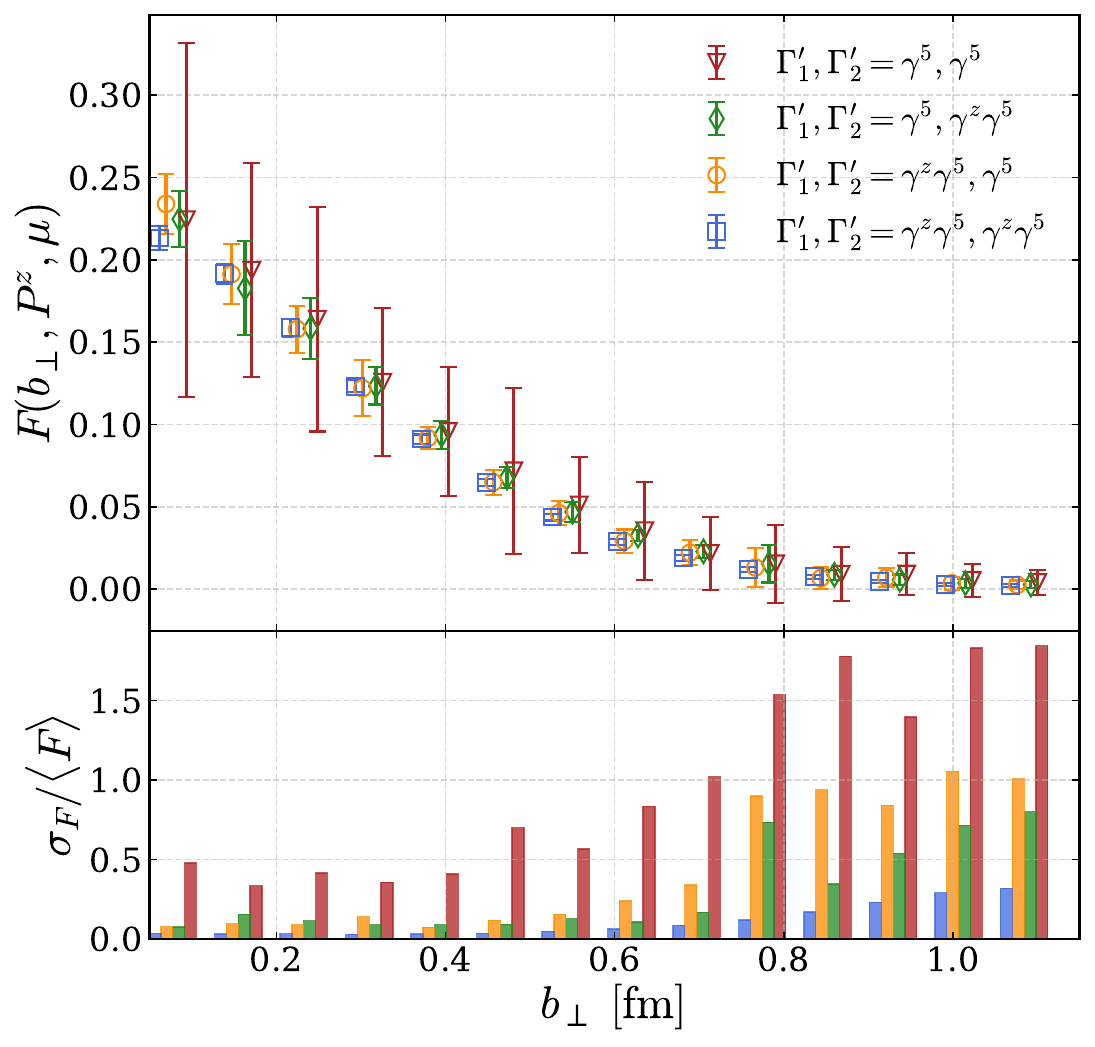}
\caption{
Renormalized form factor $F(b_\perp, P^z, \mu)$ dependence on the source and sink interpolating operators for the F32P30 ensemble. The upper panel displays a comparison of results from four operator combinations: axial-axial, axial-pseudoscalar, pseudoscalar-axial, and pseudoscalar-pseudoscalar. The lower panel shows their respective relative statistical uncertainties, visualized as unit-centered error bars. All shown uncertainties are statistical only.
}
\label{fig:interpolator_comparison}
\end{figure}

In previous lattice determinations, the statistical precision has been limited primarily by this large-momentum form factor~\cite{LatticeParton:2020uhz,Li:2021wvl,LatticePartonLPC:2023pdv,Bollweg:2025iol}. Conventionally, pseudoscalar ($\gamma_5$) interpolators are used to create and annihilate the meson states in these correlators, and the signal-to-noise ratio of this construction decreases rapidly at large boost momentum $P^z$. This limitation can be understood through a power-counting analysis of the form factor. Because the statistical variance is governed by the squared correlation function, where the external momentum phases cancel, the leading noise channels carry zero total momentum. Consequently, the noise receives no kinematic enhancement at large $P^z$~\cite{Zhang:2025hyo}. The signal size scales with the interpolator overlap $Z_{\Gamma'}(P^z)=\langle0|O_{\Gamma'}|\pi(P^z)\rangle$. Governed by the corresponding local current matrix elements, the axial-vector overlap grows linearly with momentum, $Z_A\sim\mathcal{O}(P^z)$, whereas the pseudoscalar overlap remains momentum-independent, $Z_P\sim\mathcal{O}(\Lambda_\mathrm{QCD})$. This difference leads to the hierarchy for the three-point functions:
\begin{align}
C_3^{AA}:C_3^{AP}:C_3^{PP}\sim1:\mathcal{O}\!\left(\frac{\Lambda_\mathrm{QCD}}{P^z}\right):\mathcal{O}\!\left(\frac{\Lambda_\mathrm{QCD}^2}{(P^z)^2}\right),
\label{eq:c3_scaling}
\end{align}
where $A\in\{\gamma^t\gamma_5,\gamma^z\gamma_5\}$ and $P= \{\gamma_5\}$ denote the source and sink interpolator choices. Because the variance receives no such kinematic enhancement, this signal hierarchy directly translates into a parametric improvement of the signal-to-noise ratio at large momentum ($P^z\gg \Lambda_\mathrm{QCD}$). While the interpolator dependence strictly cancels out in the exact ground-state ratio to yield the physical form factor, the relative statistical precision is retained. We therefore employ $\gamma^z\gamma_5$ at both the source and sink, which substantially improves the form-factor precision, as confirmed numerically in Fig.~\ref{fig:interpolator_comparison}, and makes a direct determination on the physical-point ensemble feasible. Detailed definitions of the nonlocal operators, the correlation functions, and the exact ratio construction are given in the Supplemental Material~\cite{supplemental}.

\begin{table}[t]
\caption{Lattice ensemble parameters and statistics. For all ensembles, $P^z=2\pi n_z/(N_s a)$ with $n_z\in\{4,5,6\}$. $N_{\widetilde{\Psi}}$ and $N_F$ denote the numbers of quasi-TMD wave function and form-factor measurements per configuration, respectively. Underlined values indicate the $N_{\tilde\psi}$ for which the quasi-TMD wave functions are computed using the same setup as in Ref.~\cite{Tan:2025ofx}.
}
\label{tab:ensembles}
\centering
\footnotesize
\setlength{\tabcolsep}{2.5pt}
\renewcommand{\arraystretch}{1.5}
\begin{ruledtabular}
\begin{tabular}{lcccc}
Ensemble & $a$ [fm] & $N_s^3\times N_t$ & $m_\pi$ [MeV] & $N_{\rm cfg}\times(N_{\tilde{\Psi}},N_F)$ \\
\hline
C32P29 &               & $32^3\times64$ & 292.4 & $984\times(\underline{4},2)$ \\
C32P23&  0.10530 & $32^3\times64$ & 228.0 & $448\times(\underline{10},24)$ \\
C48P14 &               & $48^3\times96$ & 135.5 & $204\times(\underline{24},48)$ \\ 
\hline
F32P30 & 0.07746 & $32^3\times96$ & 303.2 & $1153\times(\underline{6},6)$ \\
G36P29 & 0.06826 & $36^3\times108$ & 295.1 & $656\times(\underline{24},24)$ \\
H48P32 & 0.05187 & $48^3\times144$ & 317.2 & $550\times(\underline{8},8)$ \\
\end{tabular}
\end{ruledtabular}
\end{table}

\textit{Lattice results.}--
In this Letter, we employ six $N_f=2+1$ gauge ensembles generated by the CLQCD Collaboration using stout-smeared clover fermions and the Symanzik-improved gauge action~\cite{CLQCD:2023sdb,CLQCD:2024yyn}. These ensembles span lattice spacings $a\simeq0.052$--$0.105$~fm and pion masses $m_\pi=135.5$--$317.2$~MeV, including an ensemble at the physical point. Their parameters and statistics are summarized in Table~\ref{tab:ensembles}. Three longitudinal momenta are employed on each ensemble to resolve the finite-momentum dependence and enable the $P^z\to\infty$ extrapolation. Compared with Ref.~\cite{Tan:2025ofx}, the newly added G36P29 ensemble provides an additional intermediate lattice spacing, improving the control of the continuum extrapolation. Its quasi-TMD wave functions are analyzed in the same way as for the other ensembles, and the results are shown in the Supplemental Material~\cite{supplemental}.

The bare form factors are extracted from ground-states fits to the Euclidean-time dependence of ratios of the three-point function $C_3^{\Gamma_2'\Gamma_1'}$ and the corresponding local two-point functions, which cancel the pion overlap factors $Z_{\Gamma'}(\vec{P})$. To determine the form factors defined in Eq.~(\ref{eq:def_form_factor}), we compute the three-point functions using the current insertions $\Gamma=\gamma_i$ and $\gamma_i\gamma_5$ ($i=x,y$). At the source and sink, we choose the leading-power axial-vector interpolators $\Gamma_1'=\Gamma_2'=\gamma^z\gamma_5$ to enhance the large-momentum signal, yielding the bare form factors $F_{\gamma_i}^{(0)}$ and $F_{\gamma_i\gamma_5}^{(0)}$. The renormalized leading-twist combination entering Eq.~\eqref{eq:FF_fac} is then
\begin{align}
F(b_\perp,P^z,\mu)= \sum_{i=x,y}
\left[\frac{Z_V^2}{Z_A^2} F_{\gamma_i}^{(0)}(b_\perp,P^z) + F_{\gamma_i\gamma_5}^{(0)}(b_\perp,P^z) \right],
\label{eq:renorm_FF}
\end{align}
where $Z_V$ and $Z_A$ are the renormalization factors of the vector and axial-vector currents, respectively. Since the form factor is normalized by axial-current matrix elements, the factor $Z_A^2$ cancels in the axial-vector channel, leaving the relative factor $Z_V^2/Z_A^2$ for the vector contribution. The renormalization factors used on each ensemble are taken from Table~II of Ref.~\cite{CLQCD:2024yyn}.

Fig.~\ref{fig:interpolator_comparison} compares the form factors extracted with different source and sink interpolators. In the upper panel, the renormalized form factors obtained with the axial--axial, axial--pseudoscalar, pseudoscalar--axial, and pseudoscalar--pseudoscalar constructions are statistically consistent over the full range of $b_\perp$, confirming that the interpolator overlaps cancel in the ground-state ratio and providing a nontrivial check of the form-factor extraction. Their statistical uncertainties, displayed in the lower panel, are quite different. The axial--axial construction with $\Gamma_1'=\Gamma_2'=\gamma^z\gamma_5$ yields the smallest relative uncertainty, the two mixed choices give intermediate precision, and the conventional pseudoscalar--pseudoscalar construction is the least precise and becomes noise dominated at large $b_\perp$. This ordering agrees with the power-counting hierarchy in Eq.~\eqref{eq:c3_scaling}. The improved signal-to-noise ratio of the axial-vector interpolators yields a precise form factor over the required range of transverse separations and enables the first direct determination of the intrinsic soft function on the physical-point ensemble.

\textit{Extraction of the intrinsic soft function.} --
After determining the renormalized form factor and quasi-TMDWFs on each ensemble, we extract the intrinsic soft function from Eq.~\eqref{eq:FF_fac} through
\begin{align}
S_I^{\rm lat}(b_\perp,\mu)=&F(b_\perp,P^z,\mu)\biggl[\int dx_1dx_2\,H(x_1,x_2,P^z,\mu)\nonumber\\
&\times\widetilde{\Psi}^*(x_1,b_\perp,-P^z)\,\widetilde{\Psi}(x_2,b_\perp,P^z)\biggr]^{-1},
\label{eq:SI_extraction}
\end{align}
where the convolution is evaluated using the one-loop hard kernel of Ref.~\cite{Deng:2022gzi}. The numerator and denominator are computed on the same bootstrap samples, so that their statistical correlations are preserved in the ratio. At leading power, this ratio divides out the short-distance hard contribution and the hadron-dependent collinear dynamics, isolating the universal intrinsic soft function up to finite-momentum power corrections.

\begin{figure}[tb]
\centering
\includegraphics[width=\columnwidth]{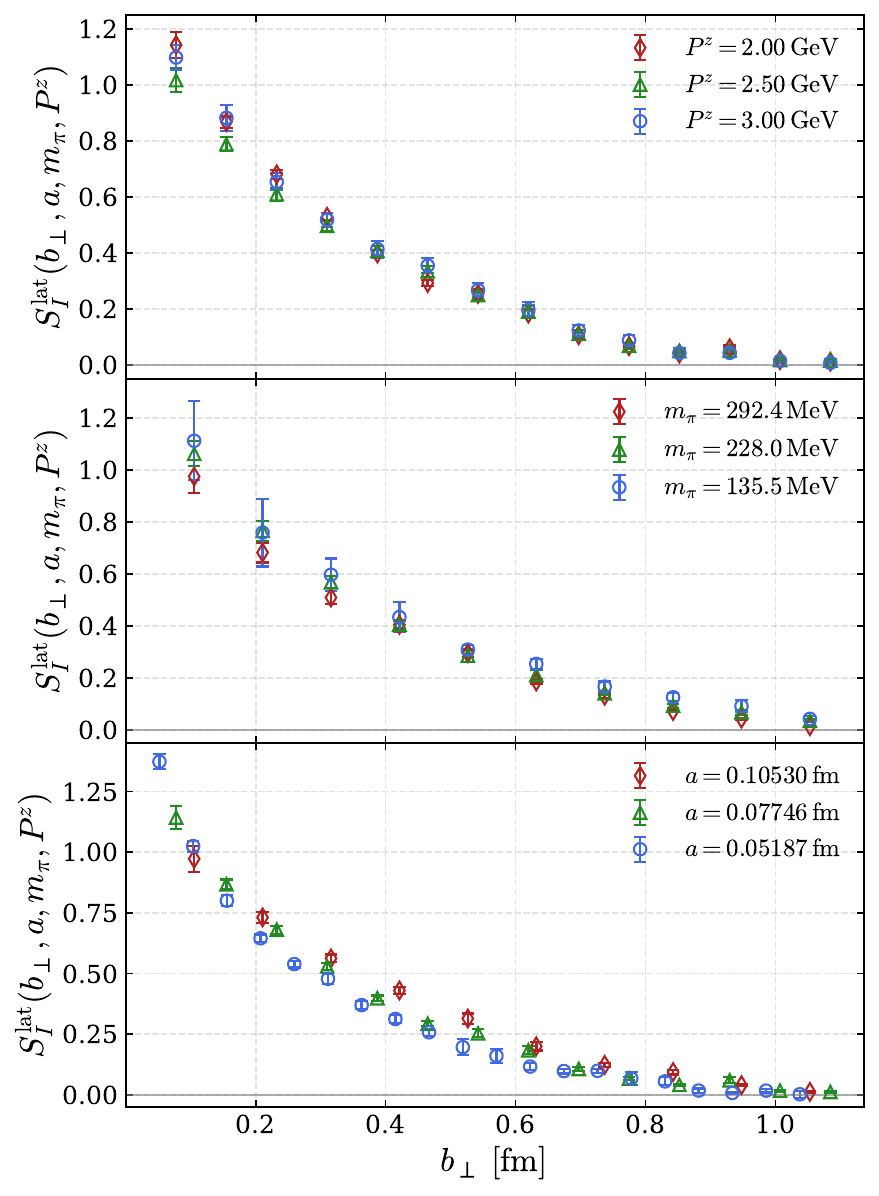}
\caption{The dependence of the extracted intrinsic soft function $S_I^{\rm lat}(b_\perp,\mu;a,m_\pi,P^z)$ on the lattice parameters is shown in the upper, middle, and lower panels, corresponding to variations in $P^z$, $m_\pi$, and $a$, respectively. The upper panel is obtained at $m_\pi=303.2$ MeV and $a=0.07746$ fm, the middle panel at $P^z=1.47$ GeV and $a=0.1053$ fm, and the lower panel at $P^z\simeq2$ GeV and $m_\pi\simeq300$ MeV. Only statistical uncertainties are shown.}
\label{fig:SI_dependence}
\end{figure}

To assess the dependence of $S_I^{\rm lat}$ on the lattice parameters, Fig.~\ref{fig:SI_dependence} shows how it varies with $P^z$, $m_\pi$, and $a$, one parameter at a time. The upper panel demonstrates that the momentum dependence in the form factor is largely canceled by the hard kernel and quasi-TMDWFs, yielding an intrinsic soft function with only weak residual $P^z$ dependence. Similarly, the middle panel shows mild sensitivity to the pion mass, including at the physical point. Since the pion mass is varied through the light-quark mass of the ensembles, this comparison probes both the external-state-mass effects in the power corrections of Eq.~\eqref{eq:FF_fac} and the sea-quark-mass dependence of $S_I$ itself. Both residual dependences are quantified in the systematic error analysis below and found to be numerically small. In contrast, the lower panel exhibits a visible lattice-spacing dependence, motivating the use of four lattice spacings for a more controllable continuum extrapolation.

\begin{figure}[tb]
\centering
\includegraphics[width=1.005\columnwidth]{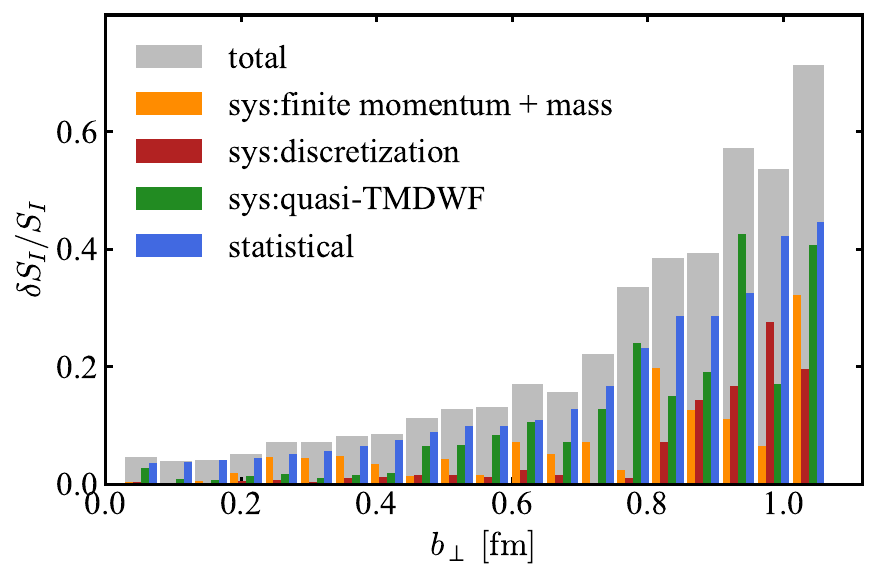}
\caption{Relative uncertainty budget for the intrinsic soft function $S_I(b_\perp,\mu)$. The orange, red, green, and blue bars denote the finite-momentum and mass uncertainty from the $P^z\to\infty$ and physical-mass extrapolations, the discretization uncertainty, the uncertainty inherited from the quasi-TMDWF analysis, and the statistical uncertainty, respectively. The gray bars show the total uncertainty obtained by adding all contributions in quadrature.}
\label{fig:syst_uncert}
\end{figure}

Guided by these observations, we obtain the intrinsic soft function $S_I(b_\perp,\mu)$ from a correlated joint fit to all ensembles and momenta at each $b_\perp$, using
\begin{align}
S_I^{\rm lat}(b_\perp,\mu&;a,m_\pi,P^z)
=S_I(b_\perp,\mu)
+\frac{A(b_\perp)}{(P^z)^2}
\nonumber\\
&
+B(b_\perp)\left[m_\pi^2-(m_\pi^{\rm phy})^2\right]+C(b_\perp)a^2 ,
\label{eq:SI_joint_extra}
\end{align}
where $A$ parameterizes the residual momentum dependence beyond the leading-power matching, and $B$ the residual pion-mass dependence. Since the pion mass is varied through the light-quark mass of the ensembles, the $B$ term captures both the residual hadron-mass effects and the sea-quark-mass dependence of $S_I$. The corresponding momentum- and mass-dependent corrections are found to be small, consistent with the expected suppression of hadron-dependent effects in the ratio of Eq.~\eqref{eq:SI_extraction}, and indicating a mild overall pion-mass dependence within the present precision. As indicated by Fig.~\ref{fig:SI_dependence}, the dominant residual dependence arises from the lattice spacing. For the continuum extrapolation, we adopt an $a^2$ dependence as the baseline ansatz, motivated by the improved lattice actions~\cite{CLQCD:2023sdb}. Possible residual linear-in-$a$ effects are assessed by including an additional term proportional to $a$ in Eq.~\eqref{eq:SI_joint_extra}, and the resulting shift in the continuum intercept is incorporated into the discretization uncertainty.

The uncertainty budget is summarized in Fig.~\ref{fig:syst_uncert}. We combine the shifts associated with the $P^z\to\infty$ and physical-mass extrapolations into a single finite-momentum and mass systematic uncertainty. To assess the continuum-extrapolation uncertainty, we include an additional term linear in $a$ and take the difference between the continuum intercepts of the $a^2$ and $a+a^2$ fits as the discretization uncertainty. The quasi-TMDWFs are Fourier transformed to momentum space after extrapolating their coordinate-space matrix elements to large $\lambda=zP^z$~\cite{Ling:2025olz,Xiong:2025obq,Chen:2025cxr}. Following Ref.~\cite{Tan:2025ofx}, we estimate the associated extrapolation uncertainty from the difference between the exponential-decay ansatz~\cite{Ji:2020brr} and the form adopted for the pion distribution amplitude in Ref.~\cite{Ji:2026vir}, and propagate it to the final result. The total uncertainty is obtained by combining all systematic contributions and the statistical uncertainty in quadrature.

\begin{figure}[tb]
\centering
\includegraphics[width=\columnwidth]{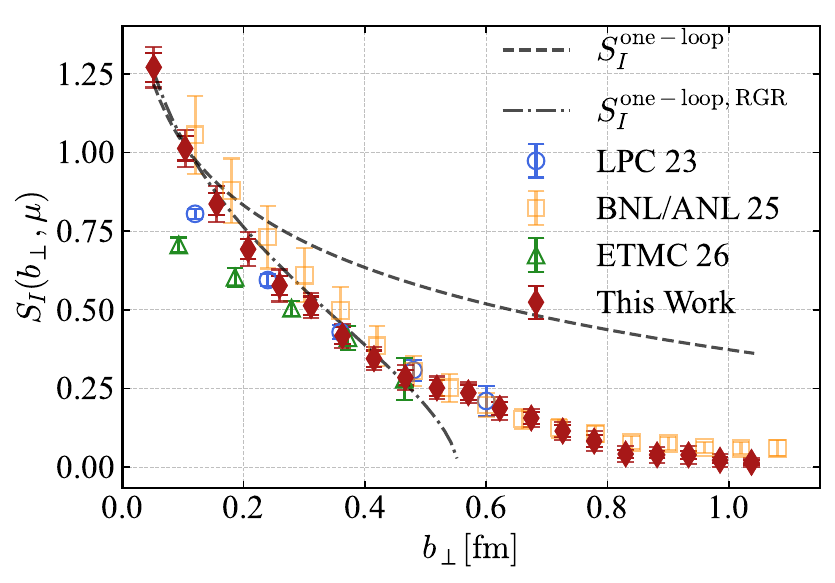}
\caption{Final intrinsic soft function at $\mu=2$~GeV, compared with previous lattice determinations~\cite{LatticePartonLPC:2023pdv,Bollweg:2025iol,Alexandrou:2025xci}, the fixed-order one-loop prediction~\cite{LatticeParton:2020uhz}, and the RGR-improved one-loop prediction~\cite{Bollweg:2025iol}. The inner error bars represent statistical uncertainties, while the outer ones show the total uncertainties obtained by adding all statistical and systematic contributions in quadrature.}
\label{fig:final_results}
\end{figure}

The final result for the intrinsic soft function, extrapolated to the continuum limit, the physical point, and infinite momentum, is presented in Fig.~\ref{fig:final_results}. After combining all statistical and systematic uncertainties, the result remains well constrained over a broad range of transverse separations. In the perturbative $b_\perp$ region, it agrees with both the fixed-order one-loop prediction~\cite{LatticeParton:2020uhz} and its renormalization-group-resummed counterpart~\cite{Bollweg:2025iol}, providing a nontrivial short-distance validation of our calculation. Toward larger $b_\perp$, where perturbation theory fails, the intrinsic soft function decreases smoothly and remains resolved up to $b_\perp\simeq1.1$~fm, deep in the nonperturbative region. Our result is broadly consistent with previous lattice determinations over their overlapping ranges~\cite{LatticePartonLPC:2023pdv,Bollweg:2025iol,Alexandrou:2025xci}. Reference~\cite{Bollweg:2025iol} employs a different extraction strategy, and the agreement with our result provides an independent cross-check of the intrinsic soft function. Although their result is obtained at an unphysical pion mass, our multi-mass analysis demonstrates that the pion-mass dependence of the intrinsic soft function is mild over the range studied, supporting the consistency of the comparison. Unlike these earlier studies, the present calculation controls all of these limits simultaneously, providing a systematically improved determination in the nonperturbative region.

\begin{figure}[tb]
\centering
\includegraphics[width=\columnwidth]{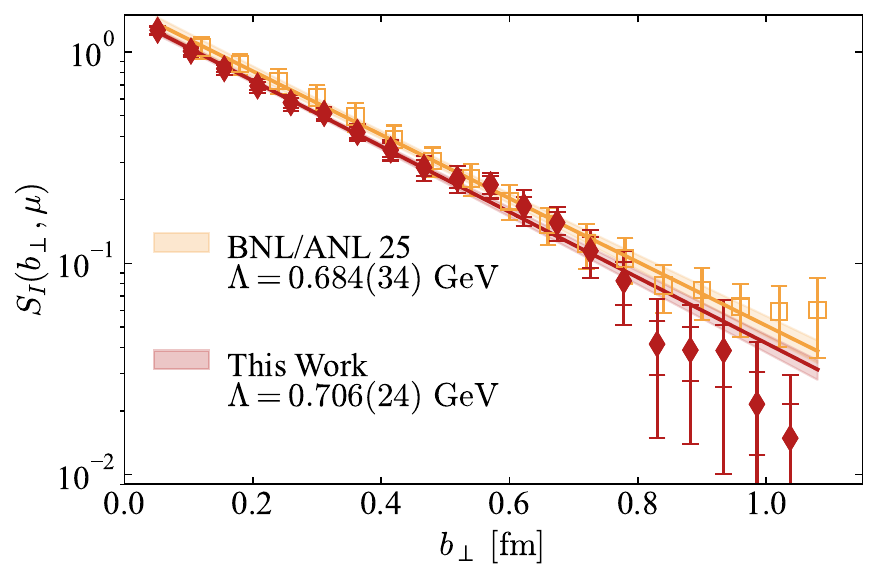}
\caption{Fits of Eq.~(\ref{eq:SI_exp}) to the intrinsic soft function at $\mu=2$~GeV. The points show the results of this work and Ref.~\cite{Bollweg:2025iol}, and the bands show the corresponding fit results with their uncertainties.}
\label{fig:decay_results}
\end{figure}

At $\mu=2~\mathrm{GeV}$, the intrinsic soft function exhibits an approximately exponential decrease with $b_\perp$, extending well into the nonperturbative region. To quantify this behavior, we fit our results over the available range of $b_\perp$ using 
\begin{align}
S_I(b_\perp)=S_0 e^{-\Lambda b_\perp},
\label{eq:SI_exp}
\end{align}
where $S_0$ is a normalization constant and $\Lambda$ characterizes the effective transverse decay scale. As shown in Fig.~\ref{fig:decay_results}, this simple form describes the observed behavior well, yielding $\Lambda=0.706(24)~\mathrm{GeV}$, corresponding to an effective attenuation length $1/\Lambda\simeq 0.28~\mathrm{fm}$. Applying the same ansatz to the independent lattice results of Ref.~\cite{Bollweg:2025iol} gives $\Lambda=0.684(34)~\mathrm{GeV}$, in good agreement with our result despite the different lattice setups and extraction strategies.

The observed exponential decaying behavior provides a quantitative characterization of the nonperturbative transverse dynamics encoded in the vacuum Wilson-line correlator defining $S_I$. Related Wilson-line structures also underlie the static quark-antiquark potential, while models of the QCD vacuum describe long-distance color correlations through field-strength correlators with finite correlation lengths ~\cite{Dosch:1994ym}. Our determination thus provides a first-principles benchmark for exploring possible connections between these nonperturbative quantities. A quantitative relation to the underlying QCD vacuum correlations or the static potential remains to be established.

\textit{Summary.}---
In this Letter, we present the first lattice QCD determination of the intrinsic soft function simultaneously extrapolated to the continuum, physical point, and infinite-momentum limits. Guided by power counting, we substantially improve the signal, enabling a direct calculation on a physical-point ensemble. Within the present precision, the extracted soft function exhibits no significant dependence on the hadron momentum, as expected from its universal vacuum origin, or on the light-quark mass, indicating that sea-quark-mass effects are small, while the visible $a$ dependence is controlled through a continuum extrapolation using four lattice spacings. Our result agrees with perturbative predictions at short distances and with previous lattice determinations over their overlapping ranges, while extending a controlled first-principles determination to $b_\perp\simeq1.1$~fm. 
Remarkably, the intrinsic soft function exhibits an approximately exponential decaying in the nonperturbative region, characterized by an effective transverse scale $\Lambda\simeq0.7~\mathrm{GeV}$. This provides a new quantitative benchmark for understanding long-distance Wilson-line correlations in the QCD vacuum. By bridging the perturbative and nonperturbative transverse-distance regimes, our calculation supplies essential universal soft input for precision studies of TMDs and three-dimensional hadron structure.

\textit{Acknowledgments.}--
We thank the CLQCD Collaboration for providing the dynamical-fermion gauge ensembles~\cite{CLQCD:2023sdb,CLQCD:2024yyn}.
This work was supported in part by the National Natural Science Foundation of China under Grant No. 12375069, 12575084, 12375080, 12125503 and 12305103 by the Fundamental Research Funds for the Central Universities.
M.-H.~C. is also supported by the National Science Centre (Poland) grant OPUS No.2021/43/B/ST2/00497. QAZ is also supported by the Fundamental Research Funds for the Central Universities. JH is also supported by Natural Science Foundation of Guangdong Province under Grant No.2025A151501219. JXT is also supported by T.D. Lee scholarship. JHZ is also supported in part by the Ministry of Science and Technology of China under Grant No. 2024YFA1611004, and by CUHK-Shenzhen under grant No. UDF01002851.
The computations in this paper were run on the Siyuan-1 cluster supported by the Center for High Performance Computing at Shanghai Jiao Tong University, and Advanced Computing East China Sub-center. The lattice QCD calculations employed the PyQUDA software suite~\cite{Jiang:2024lto} and QUDA~\cite{Clark:2009wm,Babich:2011np,Clark:2016rdz} through the HIP programming model~\cite{Bi:2020wpt}.
We have made a comparison with  LQCDMaster~\cite{Gao:2026cpd}, PhysMaster~\cite{Miao:2025sms,Tan:2026ier}, and LaMETAgent~\cite{He:2026kft} in preliminary code validation, and  data analysis during the early stages of this work.


\clearpage

\section{Supplemental Material} 

\subsection{\label{sec:enhanced}A. Signal Improvement of Form Factor from Leading-Power Axial-Vector Interpolators}

As shown in the main text, the leading-power axial-vector interpolators reduce the signal loss of the form factor at large momentum. In the main text, this improvement is explained using the pion-to-vacuum matrix elements. Here we give an explanation at the quark level.

The leading momentum dependence of the three-point function used to extract the form factor comes from the trace over the quark loop, which can be estimated as $\mathrm{Tr}[\Gamma_1'(\slashed{P}_1-m_q)\Gamma(\slashed{P}_2-m_q)\Gamma_2'(\slashed{P}_2+m_q)\Gamma(\slashed{P}_1+m_q)]$, where $m_q$ is the quark mass. Here $\Gamma$ is the Dirac structure of the inserted currents ($\gamma_i$ or $\gamma_i\gamma_5$ with $i=x,y$), and $\Gamma_1'$ and $\Gamma_2'$ are the Dirac structures of the source and sink interpolators. At large momentum, the power of $P^z$ in this trace depends strongly on the choice of $\Gamma_1'$ and $\Gamma_2'$, as summarized in Table~\ref{tab:dirac-power-counting}. For each pseudoscalar interpolator, one power of $P^z$ is replaced by $\Lambda_\mathrm{QCD}$.

\begin{figure}[htbp]
\centering
\includegraphics[width=\linewidth]{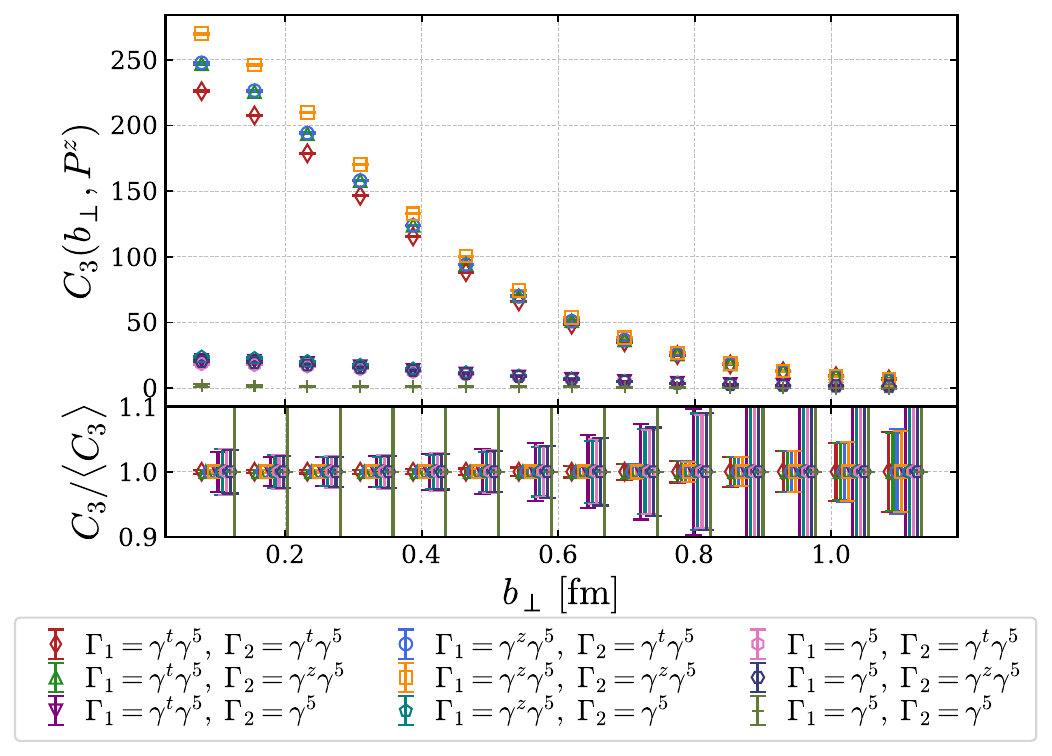}
\caption{Three-point correlation functions with different interpolators on ensemble F32P30 with $P^z=2.00$~GeV. The upper panel shows the three-point correlation functions, and the lower panel shows the corresponding signal-to-noise ratios.} 
\label{fig:snr_comparison}
\end{figure}

When the conventional pseudoscalar interpolators ($\Gamma_1'=\Gamma_2'=\gamma_5$) are used, the trace scales as $\mathcal{O}(\Lambda_\mathrm{QCD}^2(P^z)^2)$. Compared with the leading power $\mathcal{O}((P^z)^4)$, this gives a suppression factor $(\Lambda_\mathrm{QCD}/P^z)^2$ at large momentum, which corresponds to the loss of kinematic enhancement discussed in the main text. In contrast, when the axial-vector interpolators ($\gamma^t\gamma_5$ or $\gamma^z\gamma_5$) are used at both the source and the sink, the trace reaches the leading power $\mathcal{O}((P^z)^4)$. As a result, the signal does not decrease relative to the noise, which is dominated by zero-momentum states~\cite{Zhang:2025hyo}. The resulting improvement in the signal-to-noise ratio is shown in Fig.~\ref{fig:snr_comparison}. Note that for the three-point functions, this improvement mainly enhances their signal magnitudes, while the statistical uncertainties remain at a similar level across different interpolators, as shown in the upper panel of Fig.~\ref{fig:snr_comparison}. The three-point functions are normalized by two-point functions. After this normalization, the enhancement translates into a stronger form factor signal.

\begin{table}[htbp]
\centering
\renewcommand{\arraystretch}{1.35}
\setlength{\tabcolsep}{5pt}
\resizebox{\columnwidth}{!}{%
\begin{tabular}{cccc}
\hline\hline
& $\Gamma_1'=\gamma^t\gamma_5$ & $\Gamma_1'=\gamma^z\gamma_5$ & $\Gamma_1'=\gamma_5$ \\
\hline
$\Gamma_2'=\gamma^t\gamma_5$ & $\mathcal{O}((P^z)^4)$ & $\mathcal{O}((P^z)^4)$ & $\mathcal{O}(\Lambda_\mathrm{QCD}(P^z)^3)$ \\
$\Gamma_2'=\gamma^z\gamma_5$ & $\mathcal{O}((P^z)^4)$ & $\mathcal{O}((P^z)^4)$ & $\mathcal{O}(\Lambda_\mathrm{QCD}(P^z)^3)$ \\
$\Gamma_2'=\gamma_5$ & $\mathcal{O}(\Lambda_\mathrm{QCD}(P^z)^3)$ & $\mathcal{O}(\Lambda_\mathrm{QCD}(P^z)^3)$ & $\mathcal{O}(\Lambda_\mathrm{QCD}^2(P^z)^2)$ \\
\hline\hline
\end{tabular}%
}
\caption{Power counting of the trace in the three-point function at large $P^z$ for different combinations of the source ($\Gamma_1'$) and sink ($\Gamma_2'$) interpolators. }
\label{tab:dirac-power-counting}
\end{table}

\subsection{\label{sec:extraction}B. Extraction of the Form Factor and Quasi-TMDWF}

To determine the bare form factor $F_{\Gamma}^{(0)}(b_{\perp},P^z)$ and the quasi-TMD wave function $\widetilde{\Psi}(x,b_{\perp},P^z)$, we construct the corresponding three-point and two-point correlation functions. We use wall-source propagators $S_w$ in Coulomb gauge to reduce statistical noise. The initial and final meson states carry equal and opposite spatial momenta, $\vec{P}_1=-\vec{P}_2=(0,0,P^z)$, with $P^z=2\pi n_z/(N_s a)$ and $n_z\in\{4,5,6\}$. We use the momentum-projected meson interpolator
\begin{align}
O_{\Gamma'}(t,\vec{P})=\sum_{\vec{x}}e^{-i\vec{P}\cdot\vec{x}}\bar{q}_1(\vec{x},t)\Gamma'q_2(\vec{x},t).
\end{align}
The source is placed at time $0$ and the sink at time $t_{\rm seq}$. The two currents are inserted at the same time $t$ and are separated by a transverse distance $b_\perp$ perpendicular to the momentum direction. With the source interpolator $\Gamma_1'$ and the sink interpolator $\Gamma_2'$, the three-point correlation function is
\begin{align}
C&_{3}^{\Gamma_2'\Gamma_1'}(t_{\rm seq},t,b_\perp,\Gamma,P^z)=\sum_{\vec{y}}\,
\mathrm{Tr}\Big[
S^\dagger_w(t_{\rm seq},t,\vec{y};p_1)\gamma^5\Gamma_2'\nonumber\\
&\times
S_w(t_{\rm seq},t,\vec{y}+b_\perp\hat{n}_\perp;p_2)\Gamma \gamma^5S^\dagger_w(0,t,\vec{y}+b_\perp\hat{n}_\perp;p_1')\nonumber\\
&\times\gamma^5\Gamma_1'
S_w(0,t,\vec{y};p_2')\Gamma\gamma^5
\Big],
\end{align}
where $p_1$ and $p_2$ ($p_1'$ and $p_2'$) are the momenta of the wall-source propagators from the sink (source), chosen such that the final (initial) meson state has momentum $\vec{P}_2$ ($\vec{P}_1$). For $0\ll t\ll t_{\rm seq}$, the ground state dominates, and
\begin{align}
C_3^{\Gamma_2'\Gamma_1'}&\longrightarrow\frac{Z_{\Gamma_2'}(\vec{P}_2)Z_{\Gamma_1'}^*(\vec{P}_1)}{4E_{\vec{P}_1}^2}e^{-E_{\vec{P}_1}t_{\rm seq}}\nonumber\\
&\times\langle\pi(\vec{P}_2)|[\bar{q}_2\Gamma q_1](b_\perp)[\bar{q}_1\Gamma q_2](0)|\pi(\vec{P}_1)\rangle,
\end{align}
where $Z_{\Gamma'}(\vec{P})=\langle0|O_{\Gamma'}|\pi(\vec{P})\rangle$, and we have used $E_{\vec{P}_1}=E_{\vec{P}_2}$. The bare form factor $F_{\Gamma}^{(0)}(b_\perp,P^z)$ is defined as this matrix element divided by $f_\pi^2\,\vec{P}_1\cdot\vec{P}_2$. To improve the signal-to-noise ratio at large momentum, we use the leading-power axial-vector interpolators $\Gamma_1'=\Gamma_2'=\gamma^z\gamma_5$ (see Sec.~A). The inserted currents are $\Gamma=\gamma_i$ and $\gamma_i\gamma_5$ with $i=x,y$ (see Sec.~C). The two-point correlation function is
\begin{align}
C_2^{\Gamma'}&(t,z,b_\perp, P^z,\ell,\Gamma) 
=\frac{1}{N_s^3} \sum_{\vec{y}}
\operatorname{Tr} \Big\langle S_w(0, t,\vec{y}_1;-P^z/2)\nonumber \\ 
&\times \mathcal{W}_{\sqsupset}(L_w, z, b_\perp)\Gamma    
S_w(t,0,\vec{y}_2;P^z/2)\Gamma^{\prime}\times e^{izP^z} \Big\rangle,
\end{align}
where $\mathcal{W}_{\sqsupset}(L_w,z,b_\perp)$ is the staple-shaped Wilson line with longitudinal length $L_w$, and its two ends are located at $\vec{y}_1=\vec{y}+\ell\hat{n}_z+b_\perp\hat{n}_{\perp}$ and $\vec{y}_2=\vec{y}+(\ell+z)\hat{n}_z$. Here $\ell$ is the longitudinal offset of $\vec{y}_1$ relative to $\vec{y}$. For $z=b_\perp=\ell=0$, the Wilson line reduces to unity and $C_2^{\Gamma'}$ becomes a local two-point function with the current $\Gamma$ at the sink.

The bare form factor is obtained from the ratio of the three-point function to the local two-point functions with the axial-vector current $\gamma^t\gamma_5$ at the sink,
\begin{align}
&\frac{C_3^{\Gamma_2'\Gamma_1'}(t_{\rm seq},t,b_\perp,\Gamma,P^z)}{C_2^{\Gamma_2'}(t_{\rm seq}-t,0,0,-P^z,0,\gamma^t\gamma_5)\,C_2^{\Gamma_1'}(t,0,0,P^z,0,\gamma^t\gamma_5)}\nonumber\\
&=F_{\Gamma}^{(0)}(b_\perp,P^z)\frac{1+c_1(e^{-\Delta Et}+e^{-\Delta E(t_{\rm seq}-t)})}{1+c_2(e^{-\Delta Et}+e^{-\Delta E(t_{\rm seq}-t)})},
\label{eq:sm_ratio_ff}
\end{align}
where $\Delta E$ is the energy gap between the ground state and the first excited state, and $c_1$ and $c_2$ are free parameters. The local two-point functions in the denominator cancel the overlap factors $Z_{\Gamma'}(\vec{P})$ and reduce the excited-state contributions from both the initial and final states. Since $\langle0|\bar{q}_1\gamma^t\gamma_5q_2|\pi(\vec{P})\rangle\propto f_\pi E_{\vec{P}}$, their product also supplies the factor $f_\pi^2E_{\vec{P}_1}E_{\vec{P}_2}$ in the normalization of the bare form factor.

We determine the bare form factor from a joint fit of this ratio at several source-sink separations. The limit $t_{\rm seq}\to\infty$ cannot be reached in practice, because of the finite temporal extent and the fast growth of the statistical noise. We therefore use several finite but relatively large values of $t_{\rm seq}$, and separate the ground-state and excited-state contributions in the joint fit. In our analysis, we use $t_{\rm seq}$ from about $0.6$ to $1.05$~fm, which balances statistical precision and control of excited-state contamination.

\begin{figure*}[htbp]
    \centering
    \includegraphics[width=\textwidth,height=0.9\textheight,keepaspectratio]{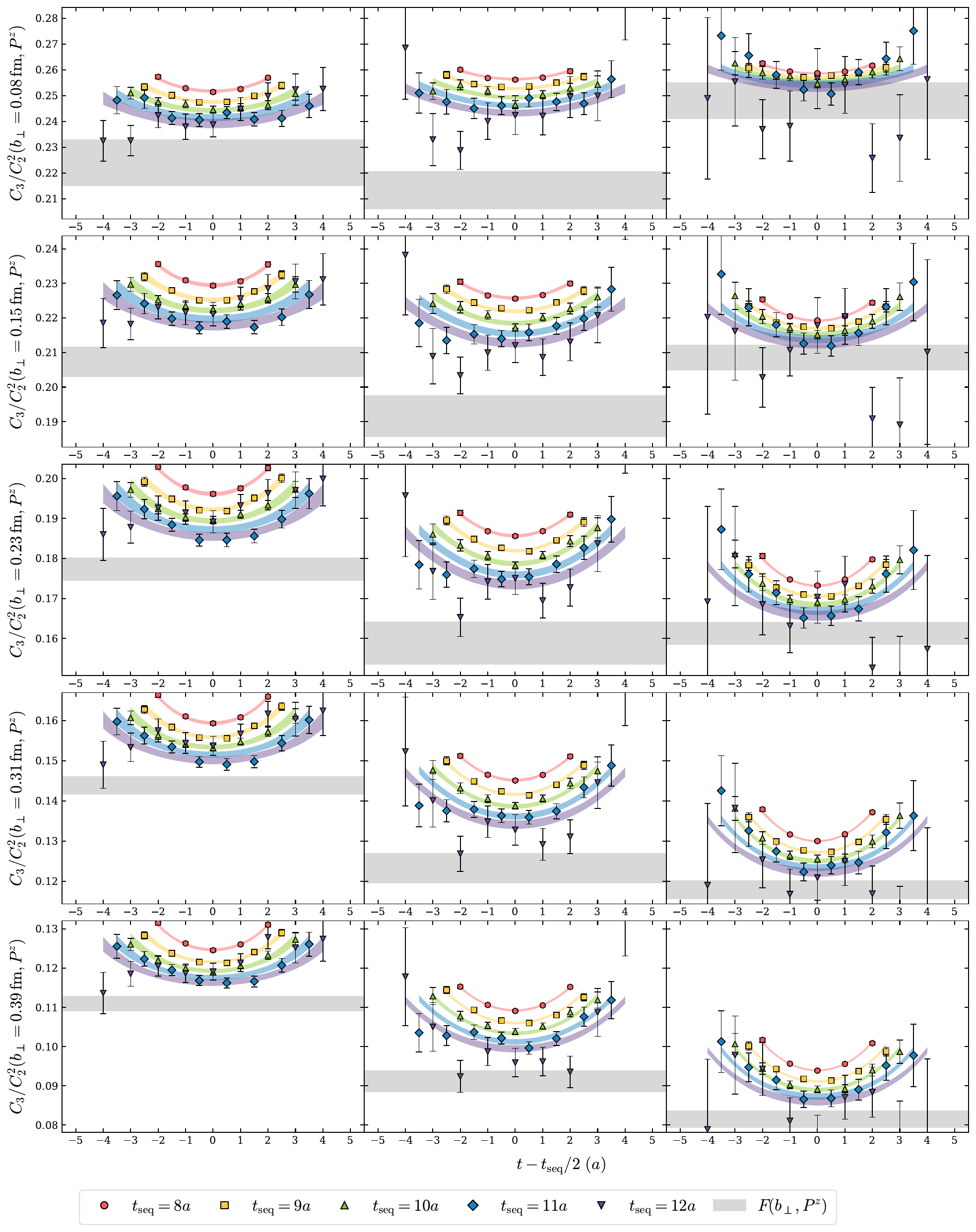}

    \caption{Joint fits of the ratio in Eq.~(\ref{eq:sm_ratio_ff}) at different source-sink separations on ensemble F32P30, for $P^z=2.00$ GeV (left), $2.50$ GeV (middle), and $3.00$ GeV (right), and $b_\perp/a=[1,5]$.}
    \label{fig:F32-ratio-to-form-factor-1-5}
\end{figure*}

\begin{figure*}[htbp]
    \centering
    \includegraphics[width=\textwidth,height=0.9\textheight,keepaspectratio]{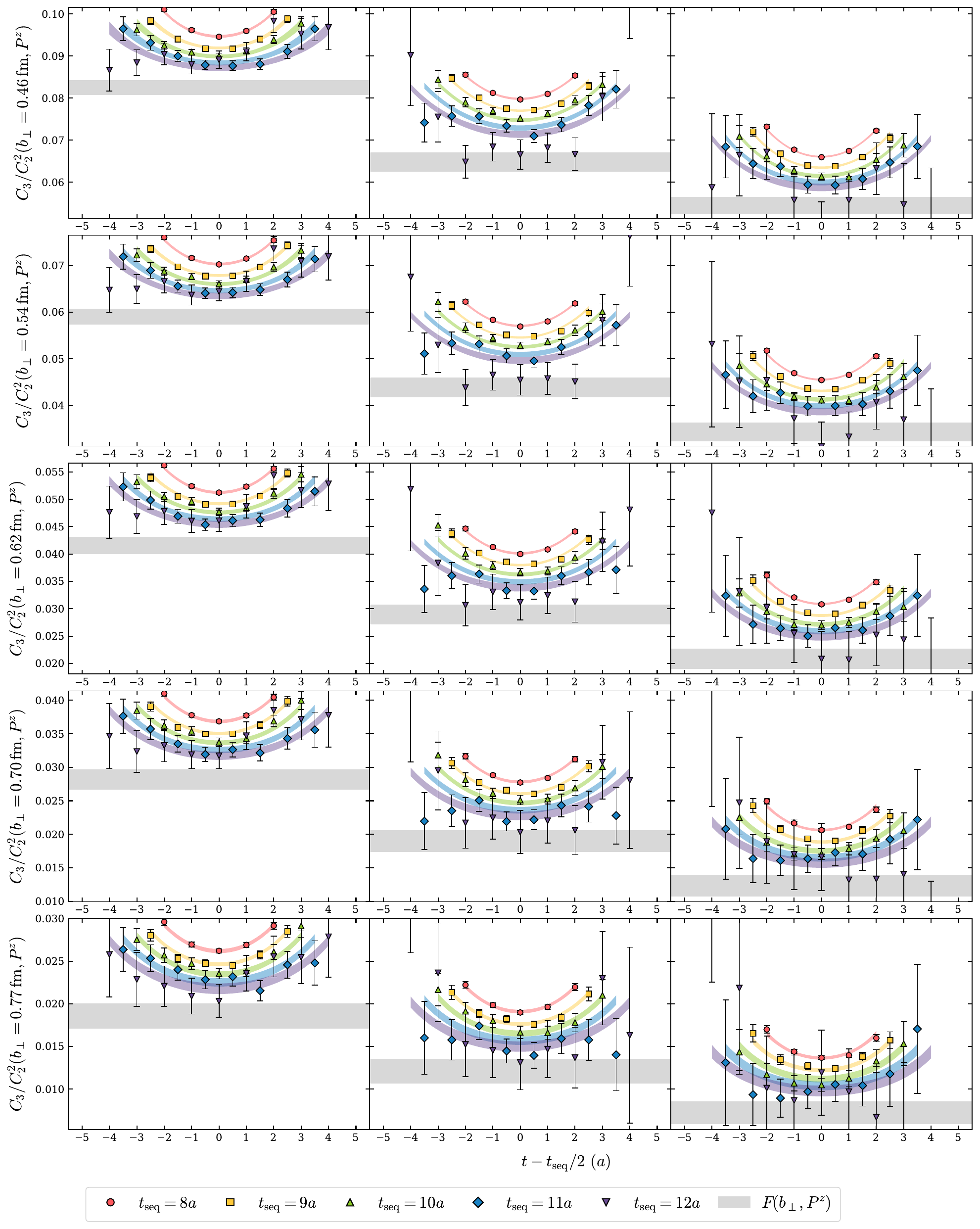}

    \caption{Joint fits of the ratio in Eq.~(\ref{eq:sm_ratio_ff}) at different source-sink separations on ensemble F32P30, for $P^z=2.00$ GeV (left), $2.50$ GeV (middle), and $3.00$ GeV (right), and $b_\perp/a=[6,10]$.}
    \label{fig:F32-ratio-to-form-factor-6-10}
\end{figure*}

\begin{figure*}[htbp]
    \centering
    \includegraphics[width=\textwidth,height=0.9\textheight,keepaspectratio]{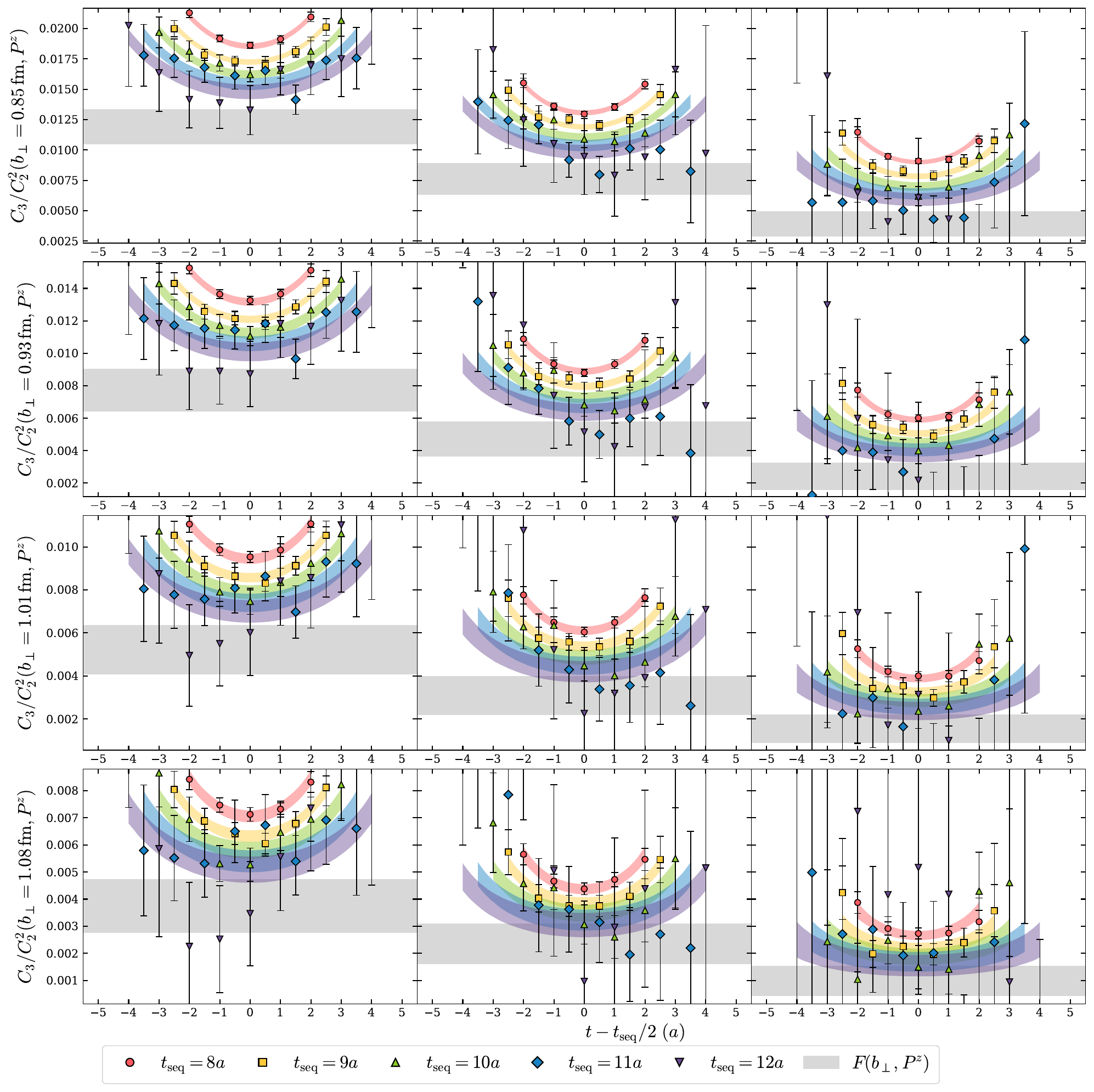}

    \caption{Joint fits of the ratio in Eq.~(\ref{eq:sm_ratio_ff}) at different source-sink separations on ensemble F32P30, for $P^z=2.00$ GeV (left), $2.50$ GeV (middle), and $3.00$ GeV (right), and $b_\perp/a=[11,14]$.}
    \label{fig:F32-ratio-to-form-factor-11-14}
\end{figure*}

\begin{figure*}[htbp]
    \centering
    \includegraphics[width=\textwidth,height=0.9\textheight,keepaspectratio]{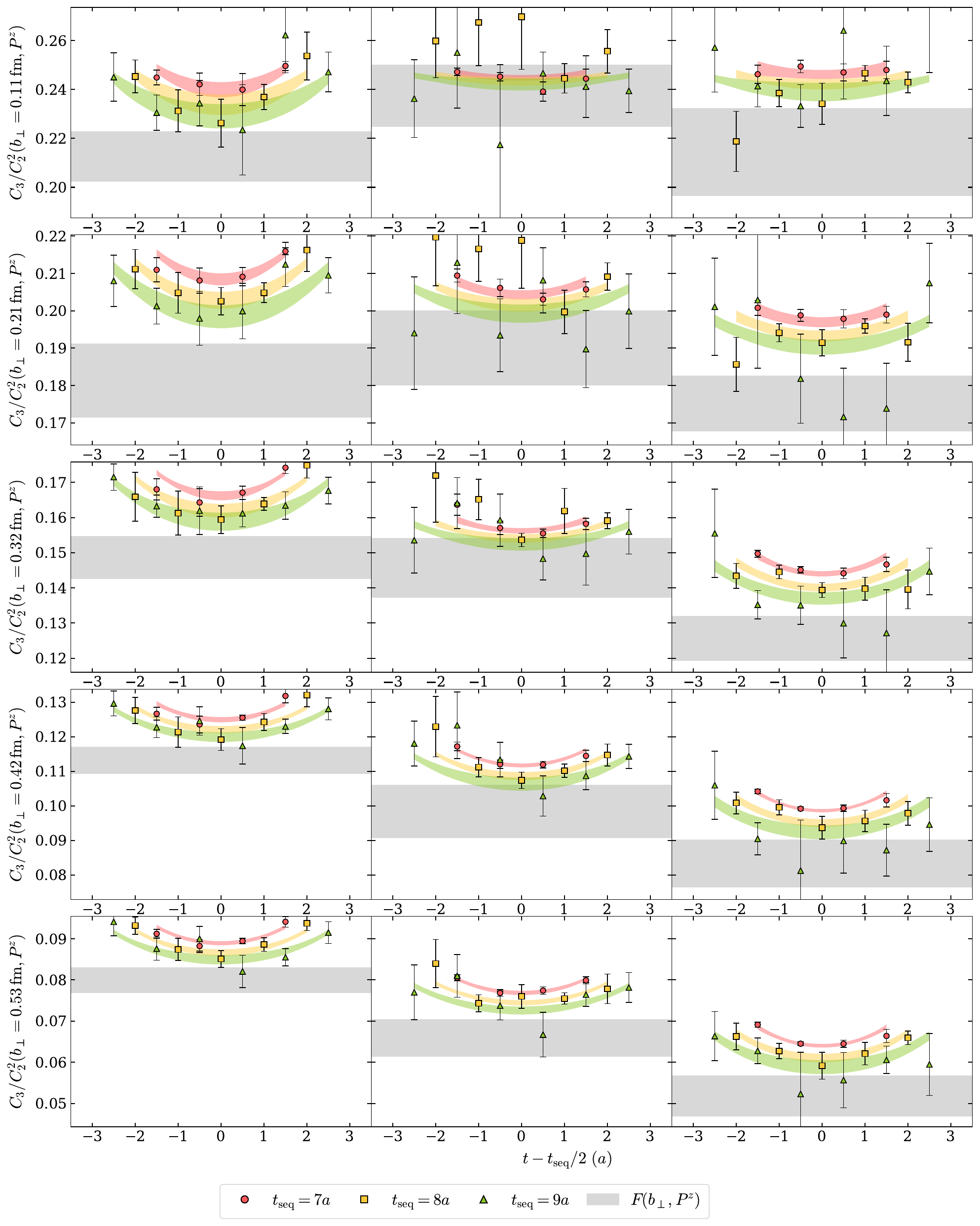}

    \caption{Joint fits of the ratio in Eq.~(\ref{eq:sm_ratio_ff}) at different source-sink separations on ensemble C32P23, for $P^z=1.47$ GeV (left), $1.84$ GeV (middle), and $2.21$ GeV (right), and $b_\perp/a=[1,5]$.}
    \label{fig:P23-ratio-to-form-factor-1-5}
\end{figure*}

\begin{figure*}[htbp]
    \centering
    \includegraphics[width=\textwidth,height=0.9\textheight,keepaspectratio]{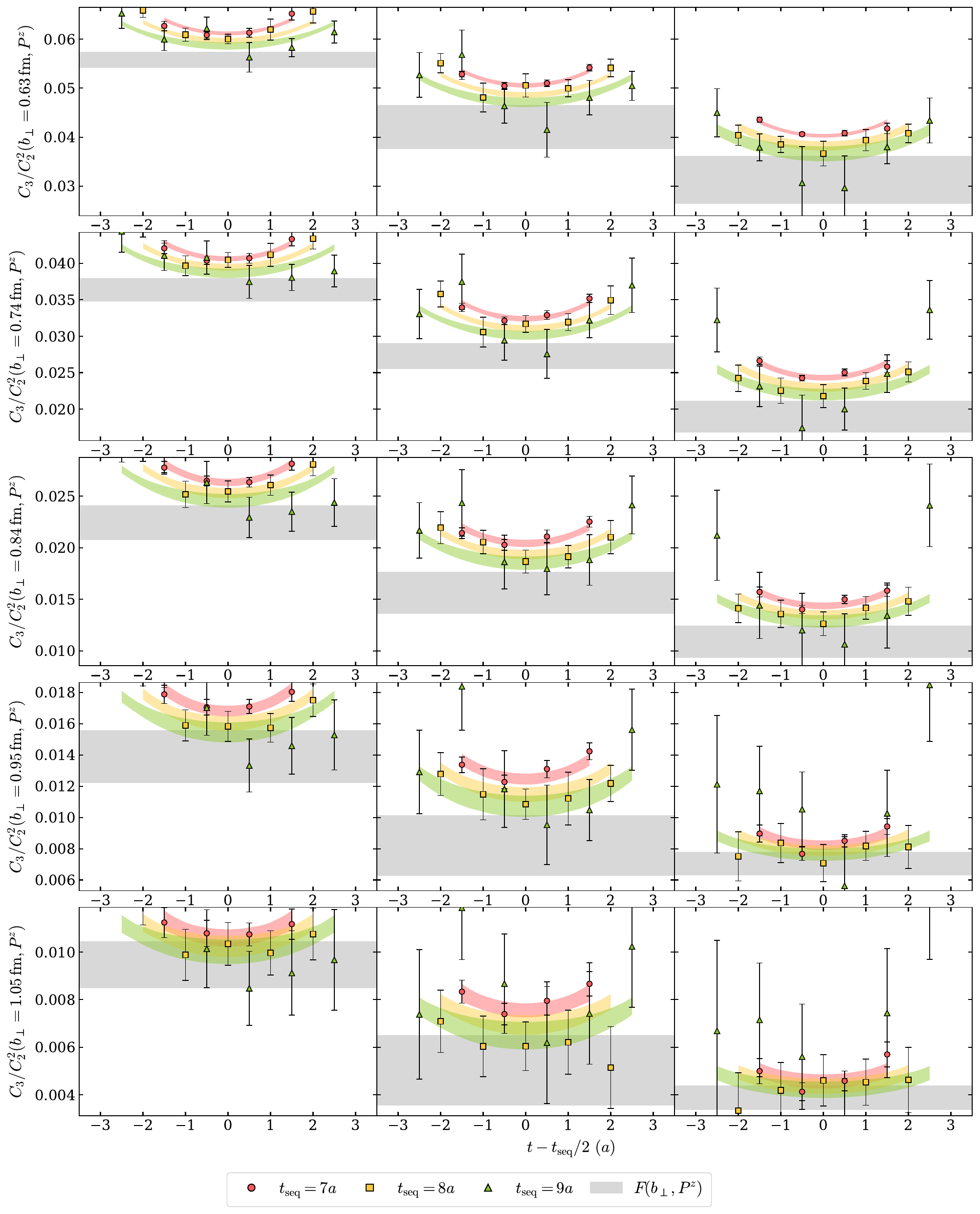}

    \caption{Joint fits of the ratio in Eq.~(\ref{eq:sm_ratio_ff}) at different source-sink separations on ensemble C32P23, for $P^z=1.47$ GeV (left), $1.84$ GeV (middle), and $2.21$ GeV (right), and $b_\perp/a=[6,10]$.}
    \label{fig:P23-ratio-to-form-factor-6-10}
\end{figure*}

\begin{figure*}[htbp]
    \centering
    \includegraphics[width=\textwidth,height=0.9\textheight,keepaspectratio]{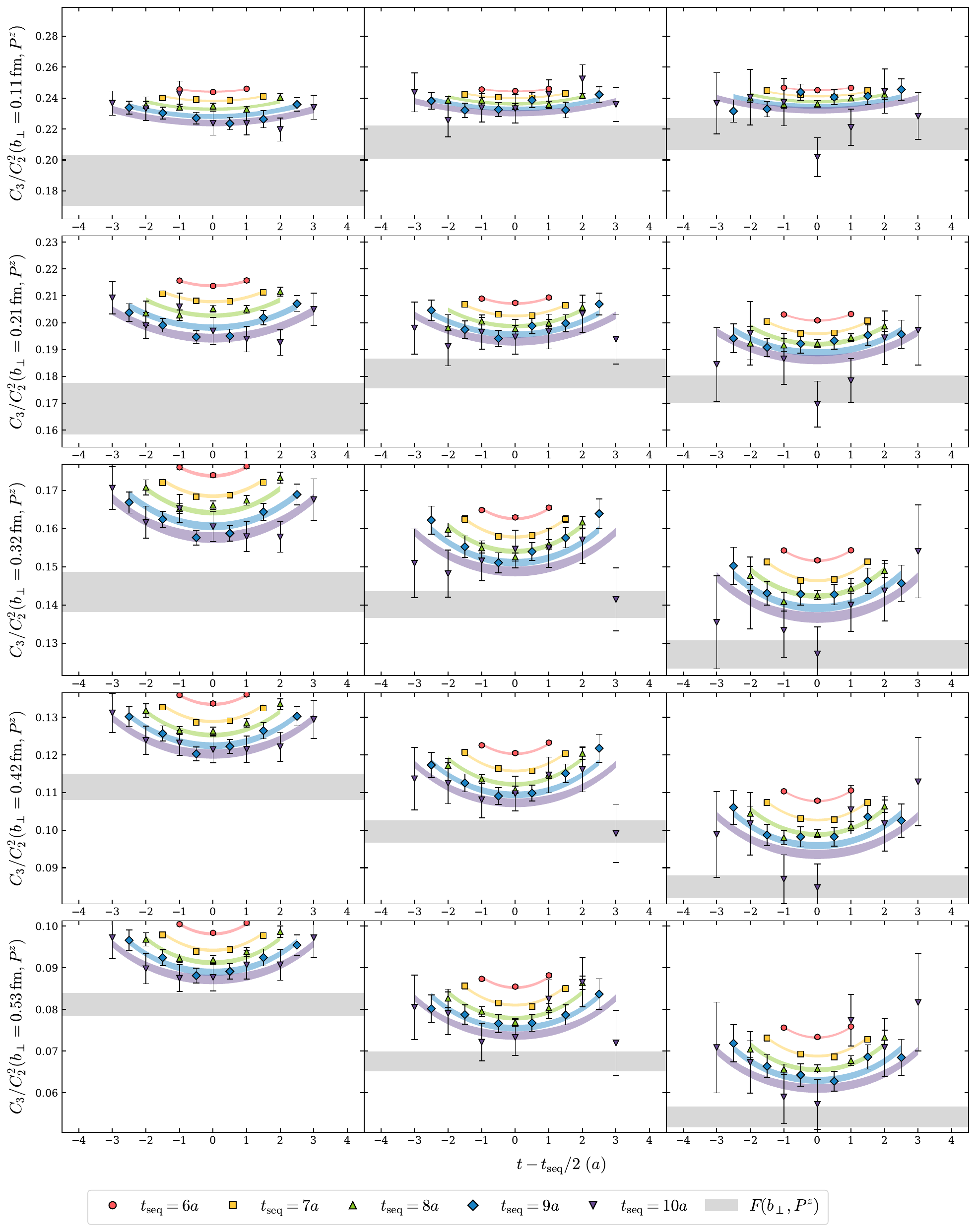}

    \caption{Joint fits of the ratio in Eq.~(\ref{eq:sm_ratio_ff}) at different source-sink separations on ensemble C32P29, for $P^z=1.47$ GeV (left), $1.84$ GeV (middle), and $2.21$ GeV (right), and $b_\perp/a=[1,5]$.}
    \label{fig:C32P29-ratio-to-form-factor-1-5}
\end{figure*}

\begin{figure*}[htbp]
    \centering
    \includegraphics[width=\textwidth,height=0.9\textheight,keepaspectratio]{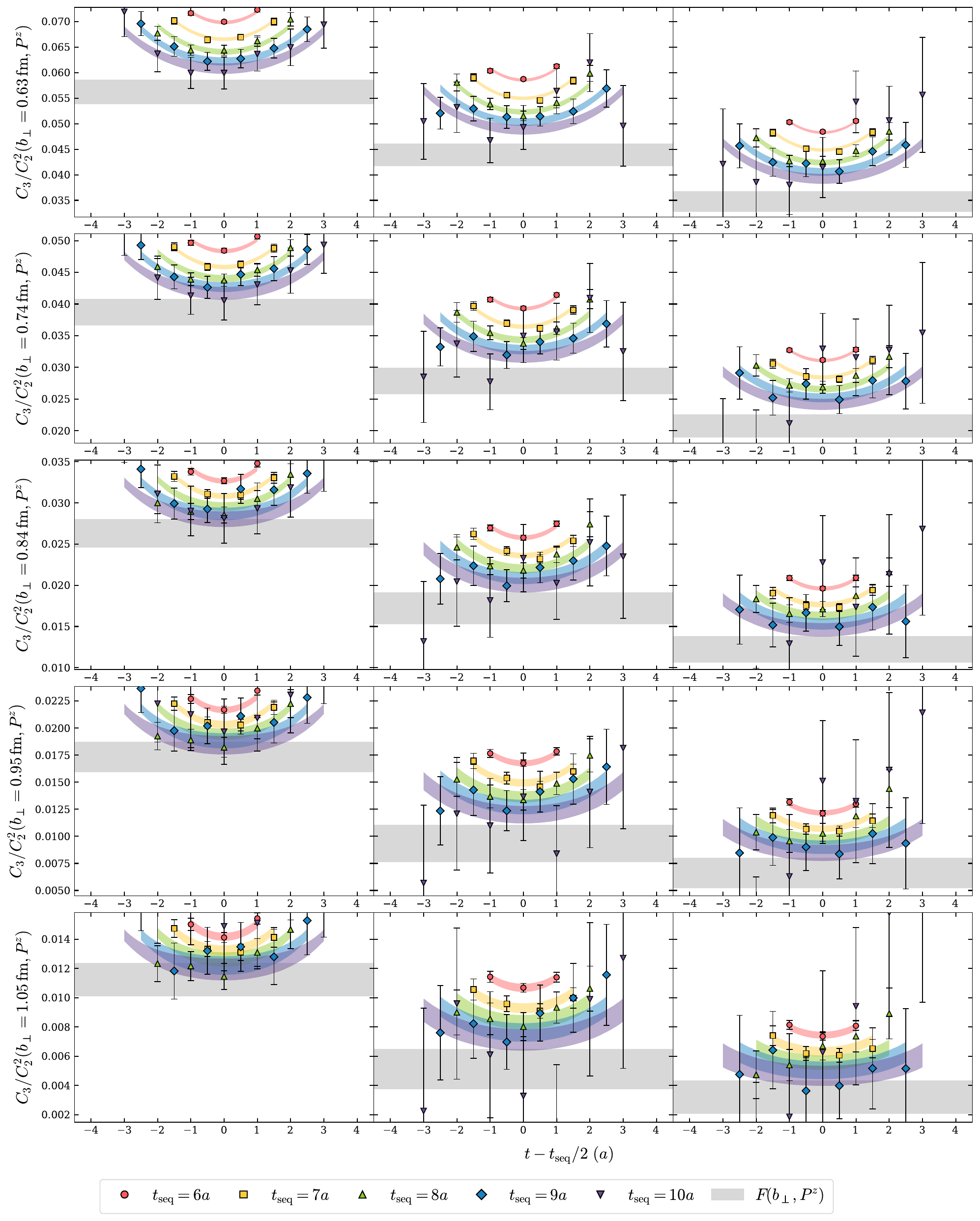}

    \caption{Joint fits of the ratio in Eq.~(\ref{eq:sm_ratio_ff}) at different source-sink separations on ensemble C32P29, for $P^z=1.47$ GeV (left), $1.84$ GeV (middle), and $2.21$ GeV (right), and $b_\perp/a=[6,10]$.}
    \label{fig:C32P29-ratio-to-form-factor-6-10}
\end{figure*}

\begin{figure*}[htbp]
    \centering
    \includegraphics[width=\textwidth,height=0.9\textheight,keepaspectratio]{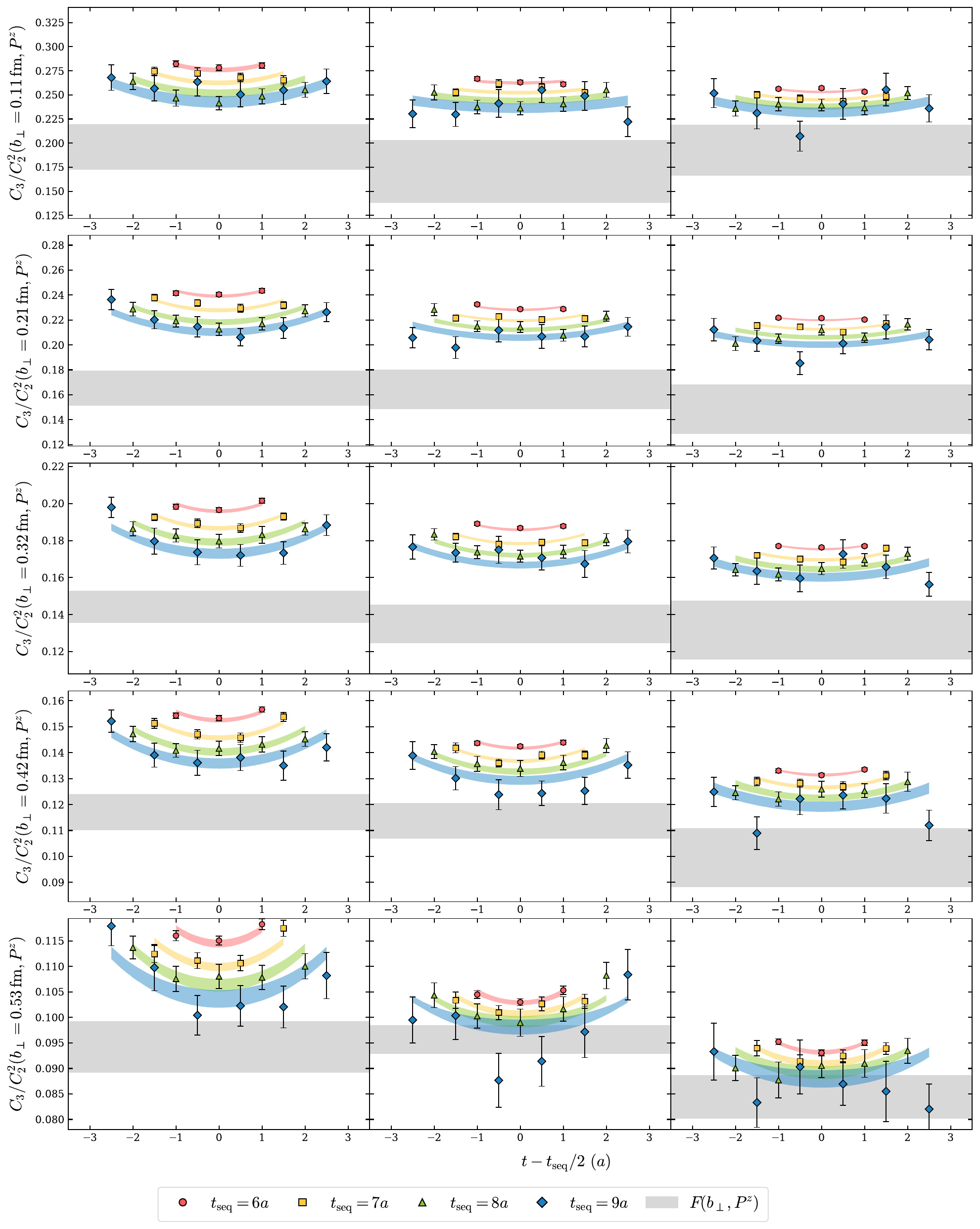}

    \caption{Joint fits of the ratio in Eq.~(\ref{eq:sm_ratio_ff}) at different source-sink separations on ensemble C48P14, for $P^z=0.98$ GeV (left), $1.23$ GeV (middle), and $1.47$ GeV (right), and $b_\perp/a=[1,5]$.}
    \label{fig:C48-ratio-to-form-factor-1-5}
\end{figure*}

\begin{figure*}[htbp]
    \centering
    \includegraphics[width=\textwidth,height=0.9\textheight,keepaspectratio]{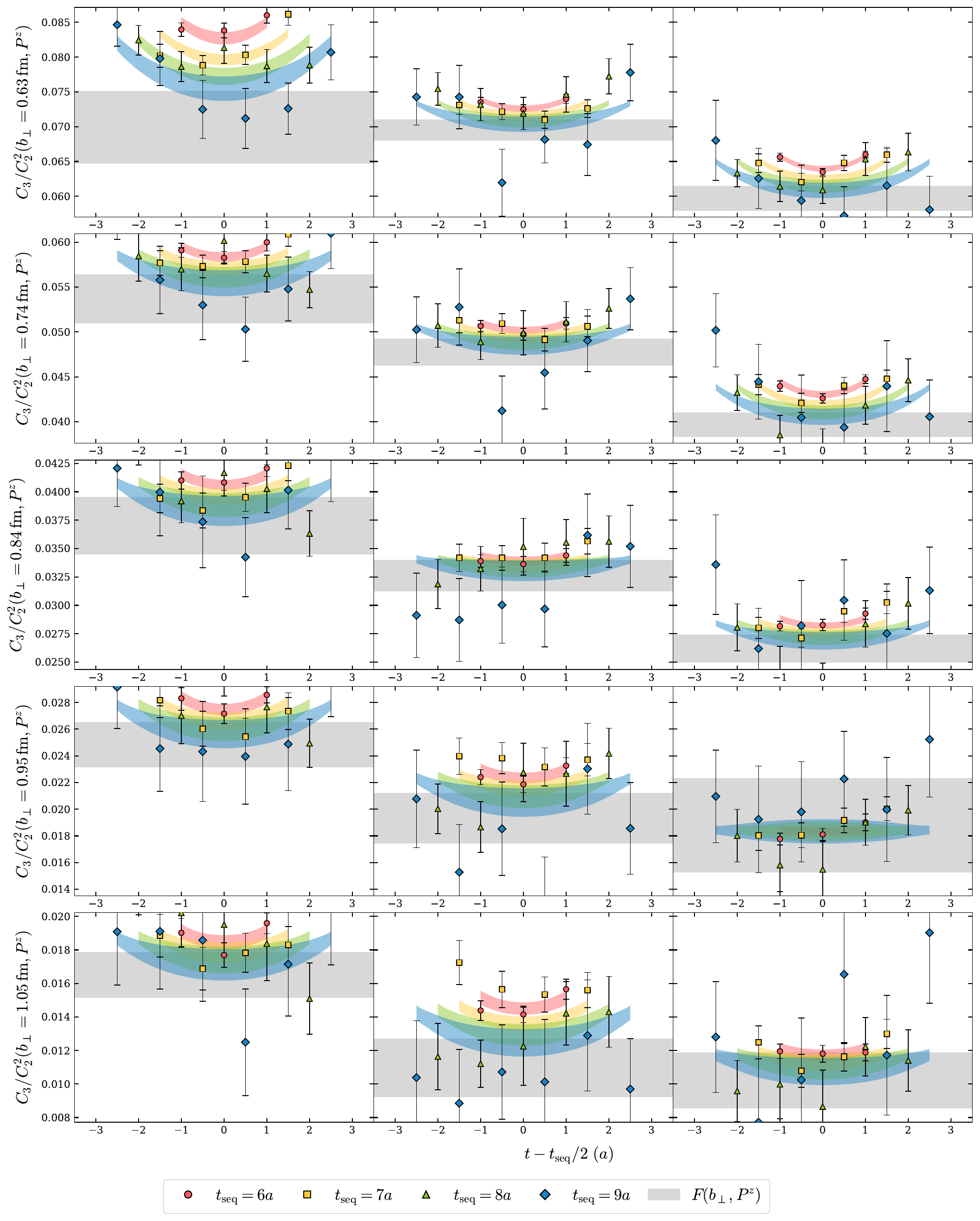}

    \caption{Joint fits of the ratio in Eq.~(\ref{eq:sm_ratio_ff}) at different source-sink separations on ensemble C48P14, for $P^z=0.98$ GeV (left), $1.23$ GeV (middle), and $1.47$ GeV (right), and $b_\perp/a=[6,10]$.}
    \label{fig:C48-ratio-to-form-factor-6-10}
\end{figure*}

\begin{figure*}[htbp]
    \centering
    \includegraphics[width=\textwidth,height=0.9\textheight,keepaspectratio]{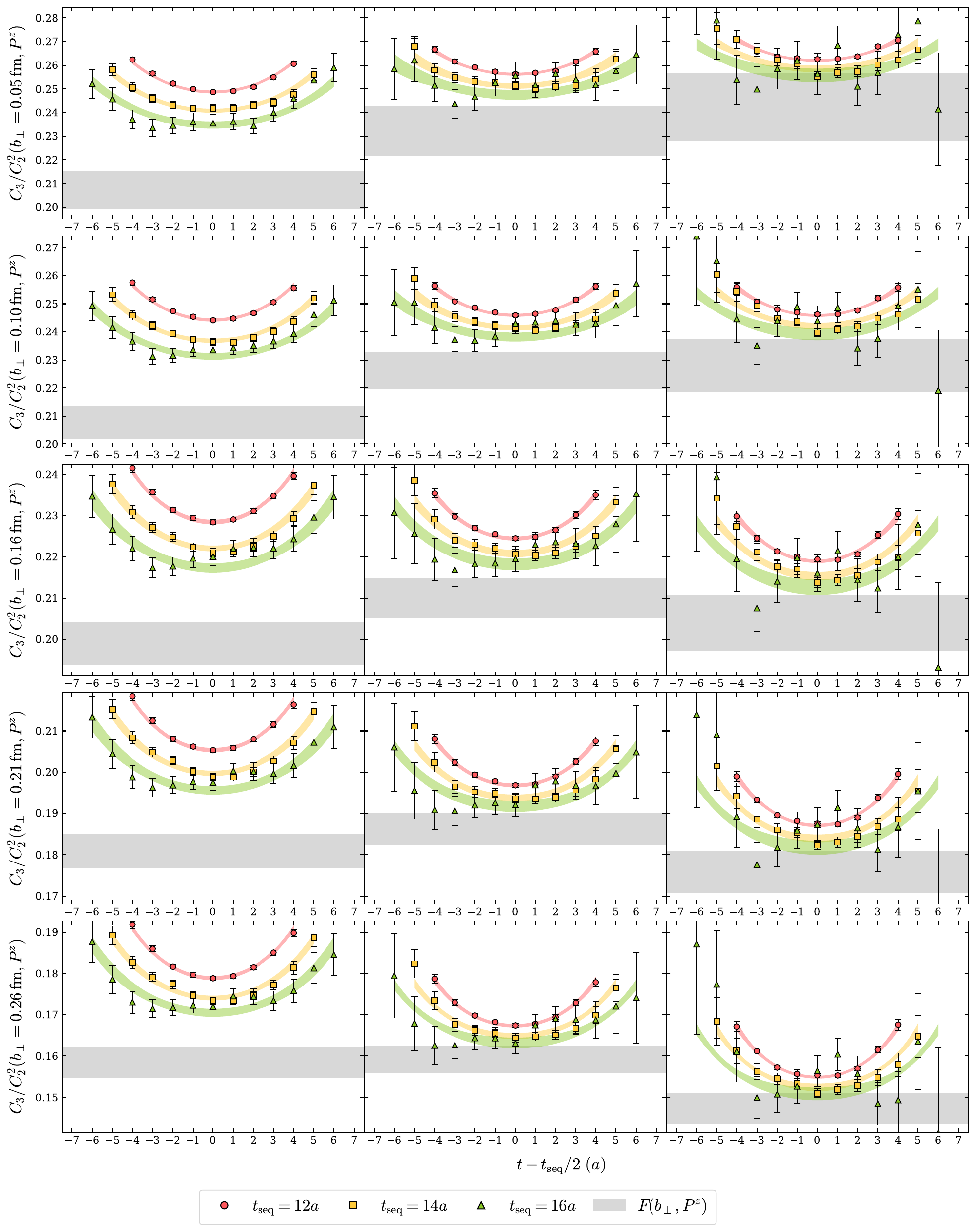}

    \caption{Joint fits of the ratio in Eq.~(\ref{eq:sm_ratio_ff}) at different source-sink separations on ensemble H48P32, for $P^z=1.99$ GeV (left), $2.49$ GeV (middle), and $2.99$ GeV (right), and $b_\perp/a=[1,5]$.}
    \label{fig:H48-ratio-to-form-factor-1-5}
\end{figure*}

\begin{figure*}[htbp]
    \centering
    \includegraphics[width=\textwidth,height=0.9\textheight,keepaspectratio]{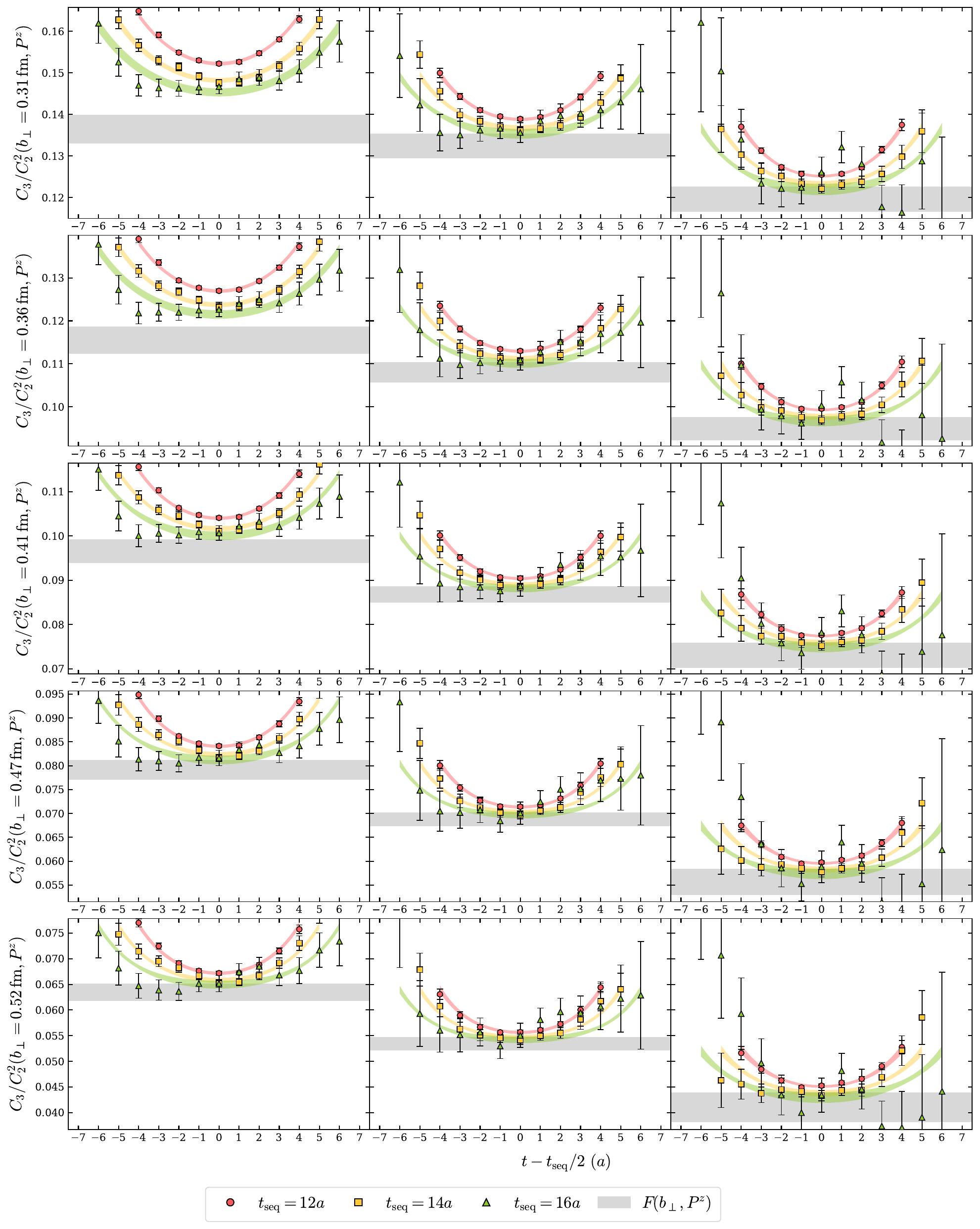}

    \caption{Joint fits of the ratio in Eq.~(\ref{eq:sm_ratio_ff}) at different source-sink separations on ensemble H48P32, for $P^z=1.99$ GeV (left), $2.49$ GeV (middle), and $2.99$ GeV (right), and $b_\perp/a=[6,10]$.}
    \label{fig:H48-ratio-to-form-factor-6-10}
\end{figure*}

\begin{figure*}[htbp]
    \centering
    \includegraphics[width=\textwidth,height=0.9\textheight,keepaspectratio]{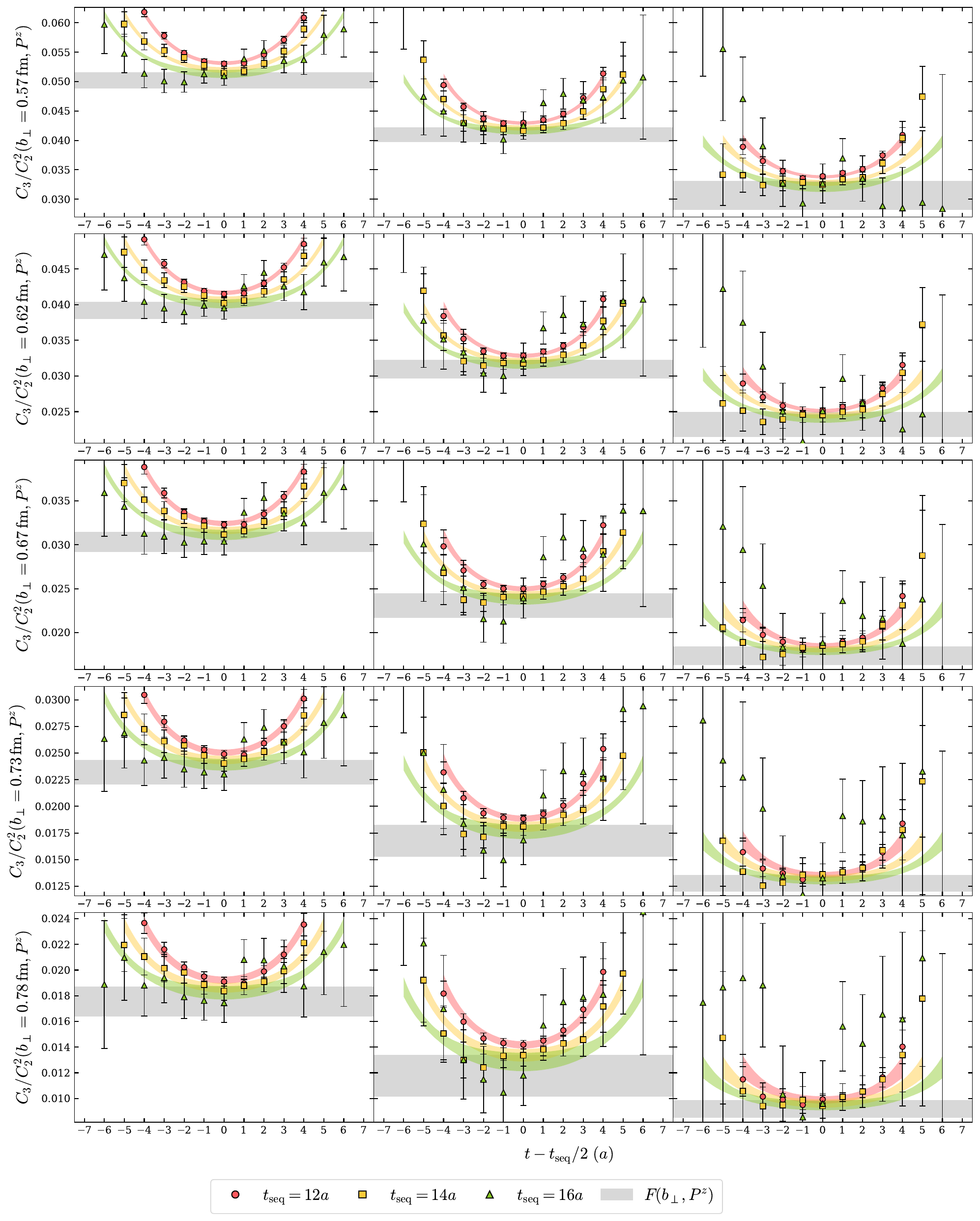}

    \caption{Joint fits of the ratio in Eq.~(\ref{eq:sm_ratio_ff}) at different source-sink separations on ensemble H48P32, for $P^z=1.99$ GeV (left), $2.49$ GeV (middle), and $2.99$ GeV (right), and $b_\perp/a=[11,15]$.}
    \label{fig:H48-ratio-to-form-factor-11-15}
\end{figure*}

\begin{figure*}[htbp]
    \centering
    \includegraphics[width=\textwidth,height=0.9\textheight,keepaspectratio]{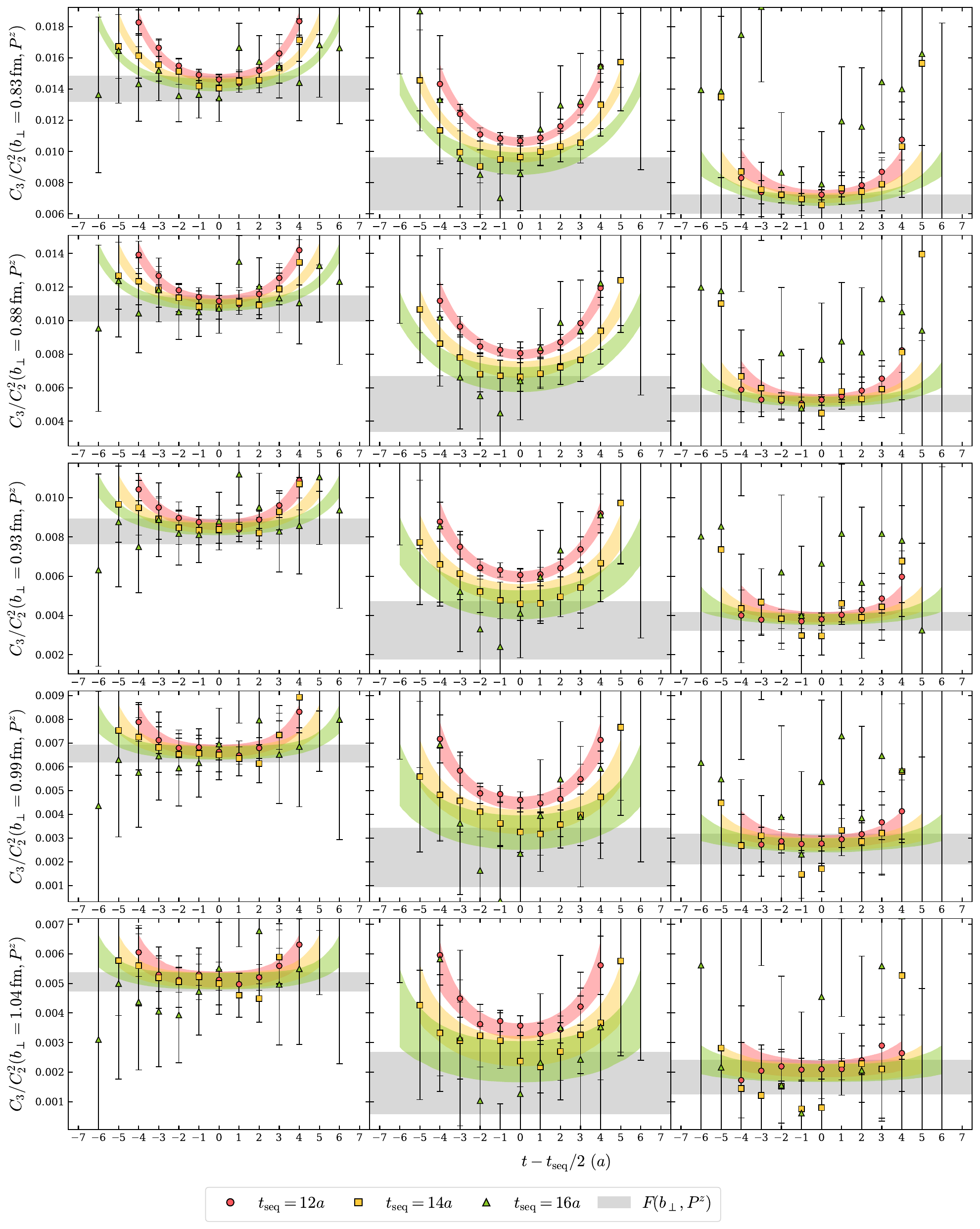}

    \caption{Joint fits of the ratio in Eq.~(\ref{eq:sm_ratio_ff}) at different source-sink separations on ensemble H48P32, for $P^z=1.99$ GeV (left), $2.49$ GeV (middle), and $2.99$ GeV (right), and $b_\perp/a=[16,20]$.}
    \label{fig:H48-ratio-to-form-factor-16-20}
\end{figure*}

\begin{figure*}[htbp]
    \centering
    \includegraphics[width=\textwidth,height=0.9\textheight,keepaspectratio]{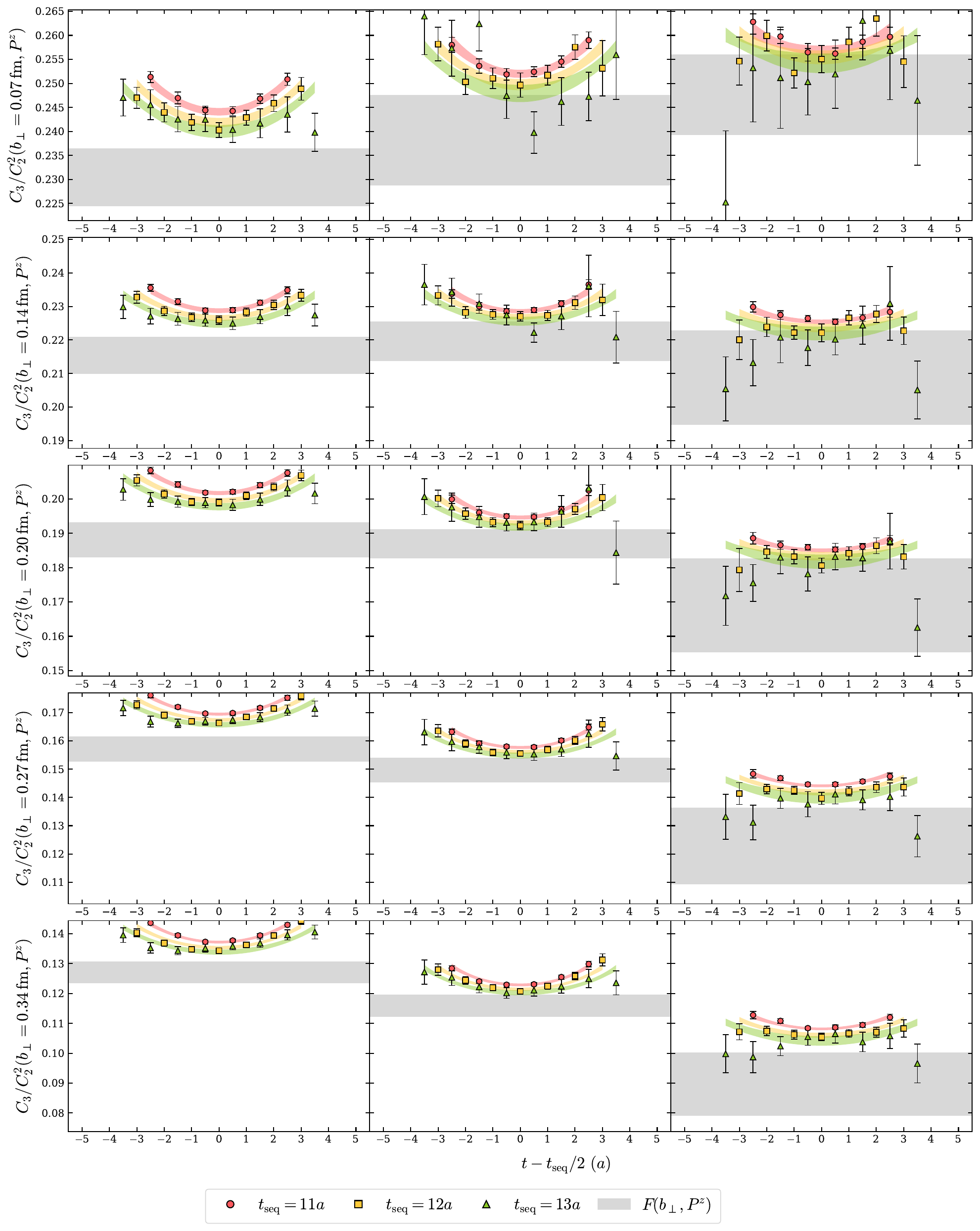}

    \caption{Joint fits of the ratio in Eq.~(\ref{eq:sm_ratio_ff}) at different source-sink separations on ensemble G36P29, for $P^z=2.02$ GeV (left), $2.52$ GeV (middle), and $3.03$ GeV (right), and $b_\perp/a=[1,5]$.}
    \label{fig:G36-ratio-to-form-factor-1-5}
\end{figure*}

\begin{figure*}[htbp]
    \centering
    \includegraphics[width=\textwidth,height=0.9\textheight,keepaspectratio]{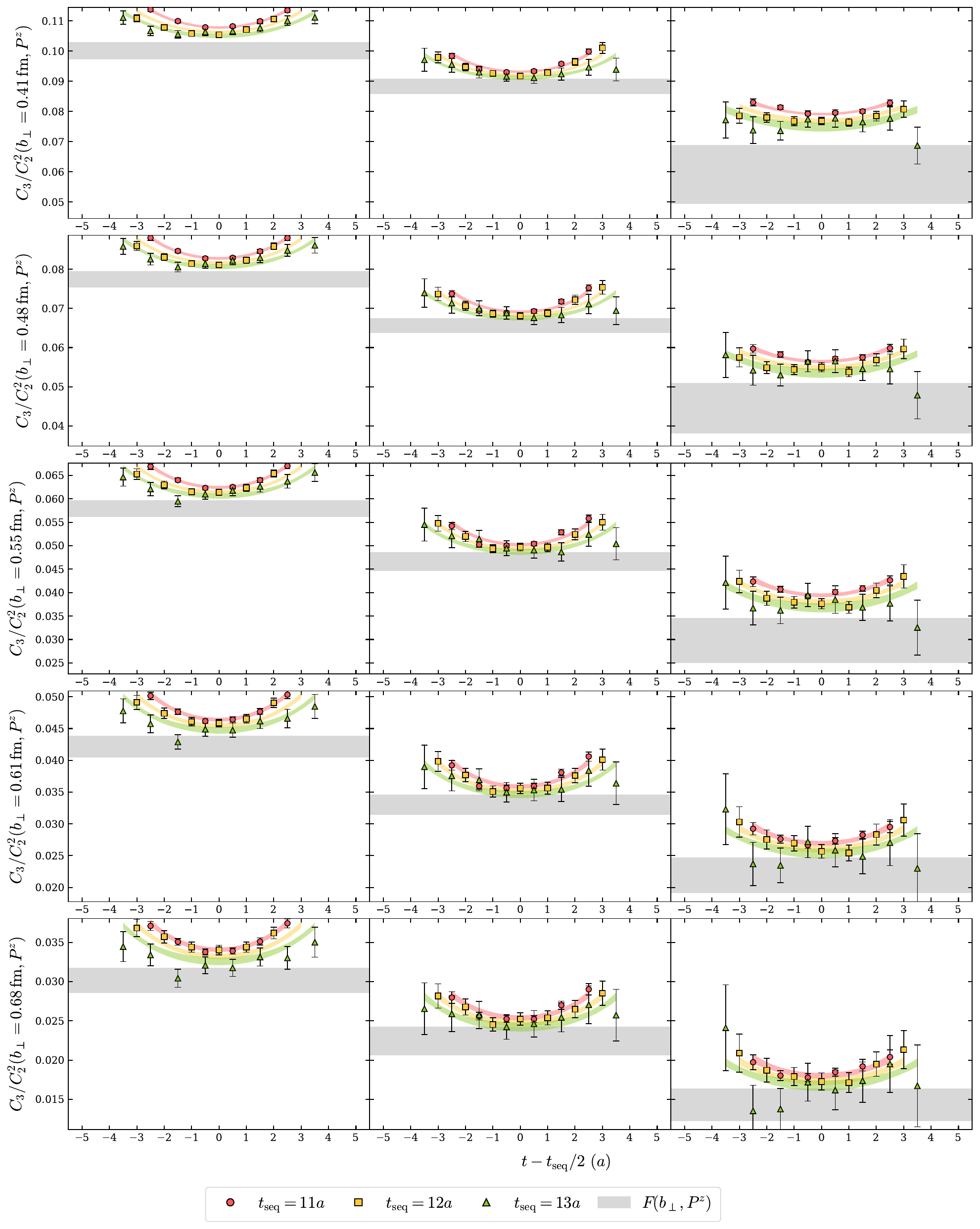}

    \caption{Joint fits of the ratio in Eq.~(\ref{eq:sm_ratio_ff}) at different source-sink separations on ensemble G36P29, for $P^z=2.02$ GeV (left), $2.52$ GeV (middle), and $3.03$ GeV (right), and $b_\perp/a=[6,10]$.}
    \label{fig:G36-ratio-to-form-factor-6-10}
\end{figure*}

\begin{figure*}[htbp]
    \centering
    \includegraphics[width=\textwidth,height=0.9\textheight,keepaspectratio]{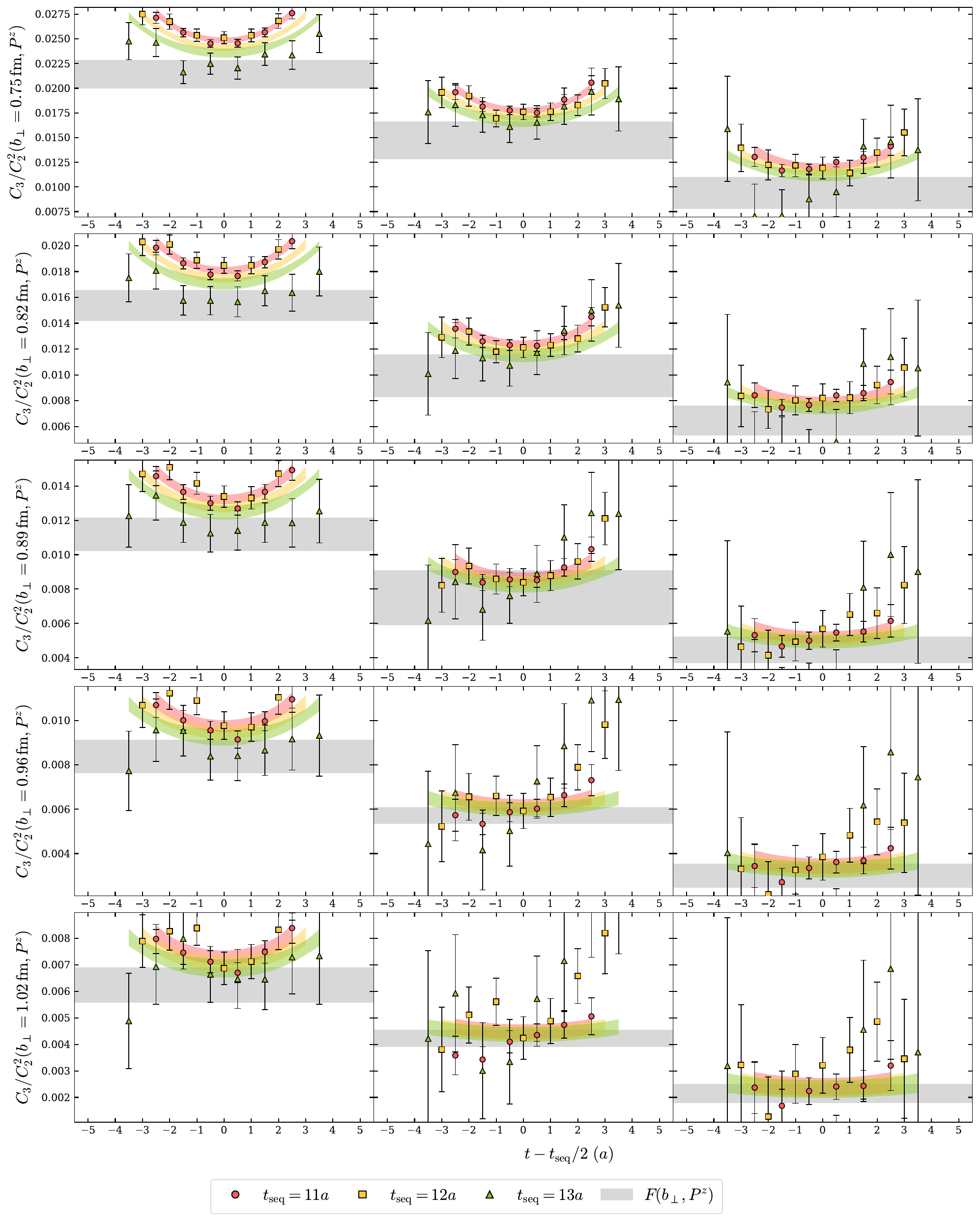}

    \caption{Joint fits of the ratio in Eq.~(\ref{eq:sm_ratio_ff}) at different source-sink separations on ensemble G36P29, for $P^z=2.02$ GeV (left), $2.52$ GeV (middle), and $3.03$ GeV (right), and $b_\perp/a=[11,15]$.}
    \label{fig:G36-ratio-to-form-factor-11-15}
\end{figure*}

Figures~\ref{fig:F32-ratio-to-form-factor-1-5} to \ref{fig:G36-ratio-to-form-factor-11-15} show the joint fits on six ensembles for $n_z=4,5,6$. The statistical uncertainty of the ratio increases quickly with $t_{\rm seq}$, as expected from the decrease of the signal-to-noise ratio. The fits are stable for all the kinematic setups shown, and the relative statistical uncertainties of the ground-state form factors are below 8\%.


\begin{figure*}[htbp]
    \centering

    \includegraphics[width=\textwidth,height=0.9\textheight,keepaspectratio]{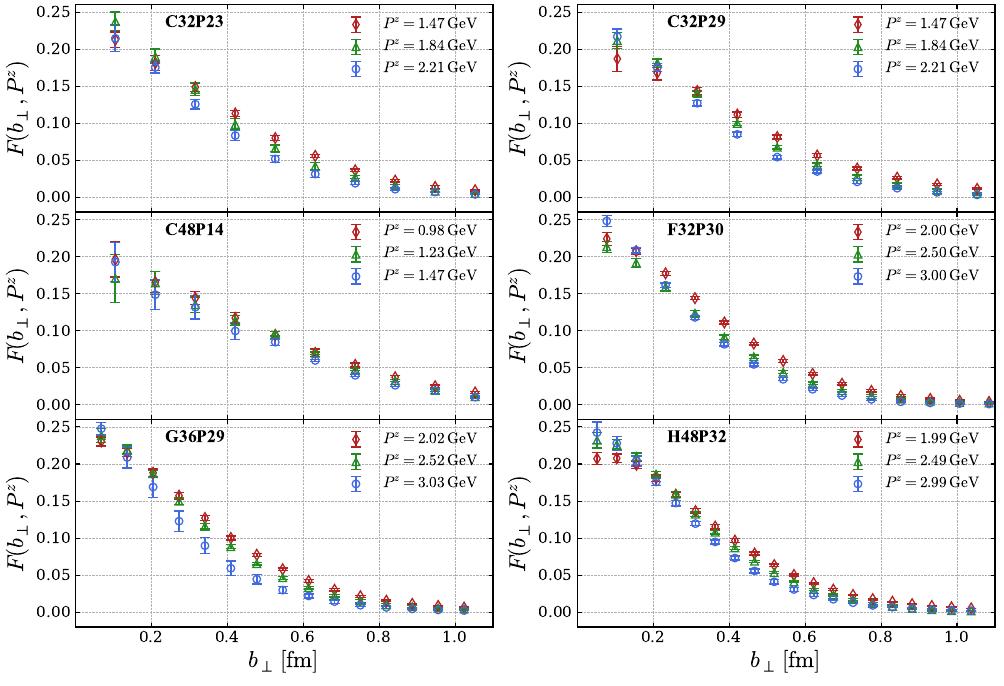}

    \caption{Form factors on different ensembles. From left to right and top to bottom, the panels correspond to C32P23, C32P29, C48P14, F32P30, G36P29, and H48P32.}
    \label{fig:form-factor-each-ensemble}
\end{figure*}

We apply the same procedure to all momenta and transverse separations on all ensembles listed in Table~I of the main text. The results are shown in Fig.~\ref{fig:form-factor-each-ensemble}. The form factors show a clear dependence on $P^z$.

\begin{figure*}[htbp]
    \centering
    \includegraphics[width=\textwidth,
    trim= 0 0.2cm 0 0,
    clip]{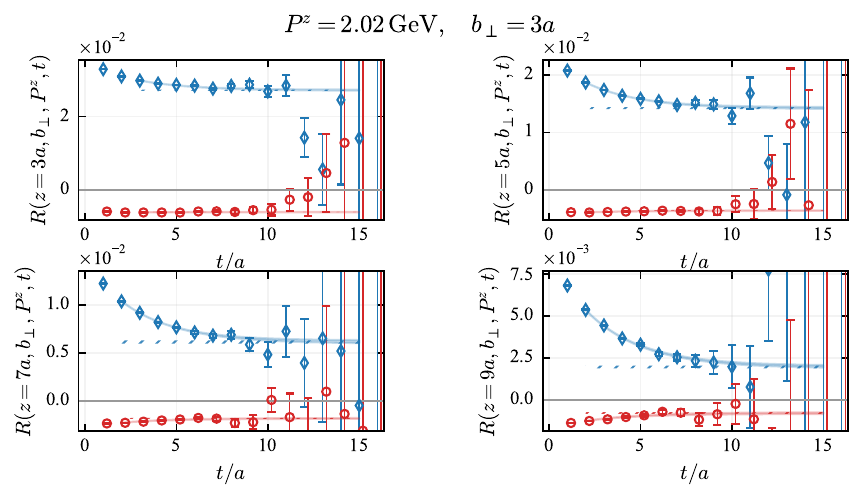}

    \caption{Extraction of the bare quasi-TMDWF matrix elements on ensemble G36P29 with $P^z=2.02$~GeV, $b_\perp=3a$, and $z=3,5,7$ and $9a$. Blue and red points denote the real and imaginary parts, and the dashed lines denote the plateau bands.}
    \label{fig:G36-bare-ME-pz4-b3-z3-9}
\end{figure*}

\begin{figure*}[htbp]
    \centering
    \includegraphics[width=\textwidth,
    trim= 0 0.2cm 0 0,
    clip]{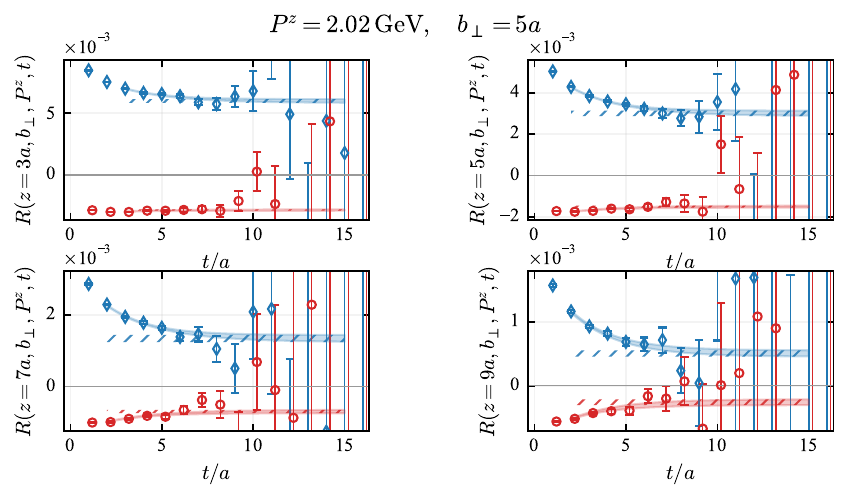}

    \caption{Extraction of the bare quasi-TMDWF matrix elements on ensemble G36P29 with $P^z=2.02$~GeV, $b_\perp=5a$, and $z=3,5,7$ and $9a$. Blue and red points denote the real and imaginary parts, and the dashed lines denote the plateau bands.}
    \label{fig:G36-bare-ME-pz4-b5-z3-9}
\end{figure*}

\begin{figure*}[htbp]
    \centering
    \includegraphics[width=\textwidth,
    trim= 0 0.2cm 0 0,
    clip]{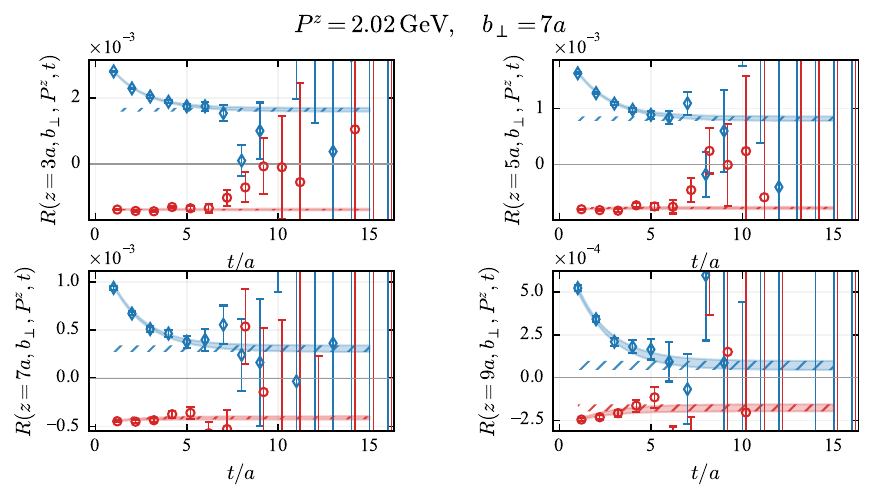}

    \caption{Extraction of the bare quasi-TMDWF matrix elements on ensemble G36P29 with $P^z=2.02$~GeV, $b_\perp=7a$, and $z=3,5,7$ and $9a$. Blue and red points denote the real and imaginary parts, and the dashed lines denote the plateau bands.}
    \label{fig:G36-bare-ME-pz4-b7-z3-9}
\end{figure*}

\begin{figure*}[htbp]
    \centering
    \includegraphics[width=\textwidth,
    trim= 0 0.2cm 0 0,
    clip]{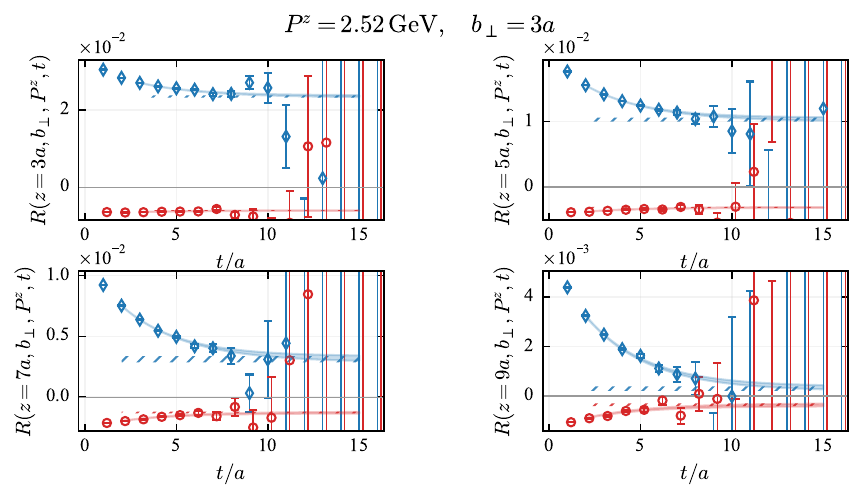}

    \caption{Extraction of the bare quasi-TMDWF matrix elements on ensemble G36P29 with $P^z=2.52$~GeV, $b_\perp=3a$, and $z=3,5,7$ and $9a$. Blue and red points denote the real and imaginary parts, and the dashed lines denote the plateau bands.}
    \label{fig:G36-bare-ME-pz5-b3-z3-9}
\end{figure*}

\begin{figure*}[htbp]
    \centering
    \includegraphics[width=\textwidth,
    trim= 0 0.2cm 0 0,
    clip]{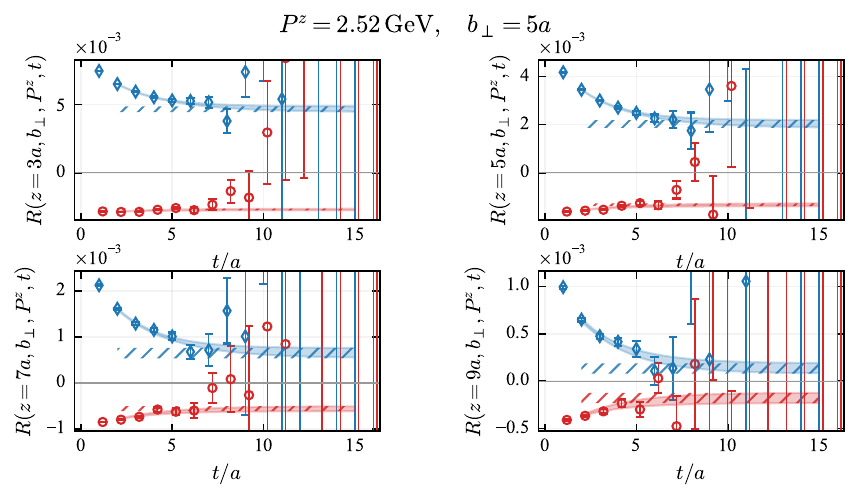}

    \caption{Extraction of the bare quasi-TMDWF matrix elements on ensemble G36P29 with $P^z=2.52$~GeV, $b_\perp=5a$, and $z=3,5,7$ and $9a$. Blue and red points denote the real and imaginary parts, and the dashed lines denote the plateau bands.}
    \label{fig:G36-bare-ME-pz5-b5-z3-9}
\end{figure*}

\begin{figure*}[htbp]
    \centering
    \includegraphics[width=\textwidth,
    trim= 0 0.2cm 0 0,
    clip]{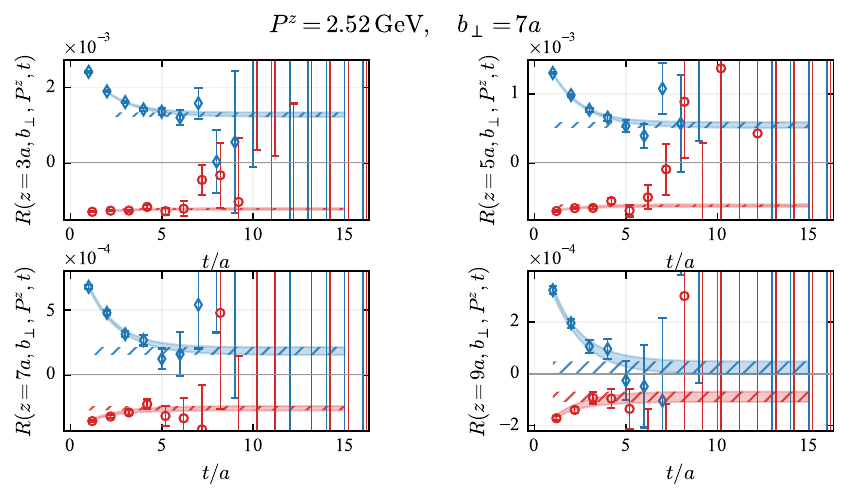}

    \caption{Extraction of the bare quasi-TMDWF matrix elements on ensemble G36P29 with $P^z=2.52$~GeV, $b_\perp=7a$, and $z=3,5,7$ and $9a$. Blue and red points denote the real and imaginary parts, and the dashed lines denote the plateau bands.}
    \label{fig:G36-bare-ME-pz5-b7-z3-9}
\end{figure*}

\begin{figure*}[htbp]
    \centering
    \includegraphics[width=\textwidth,
    trim= 0 0.2cm 0 0,
    clip]{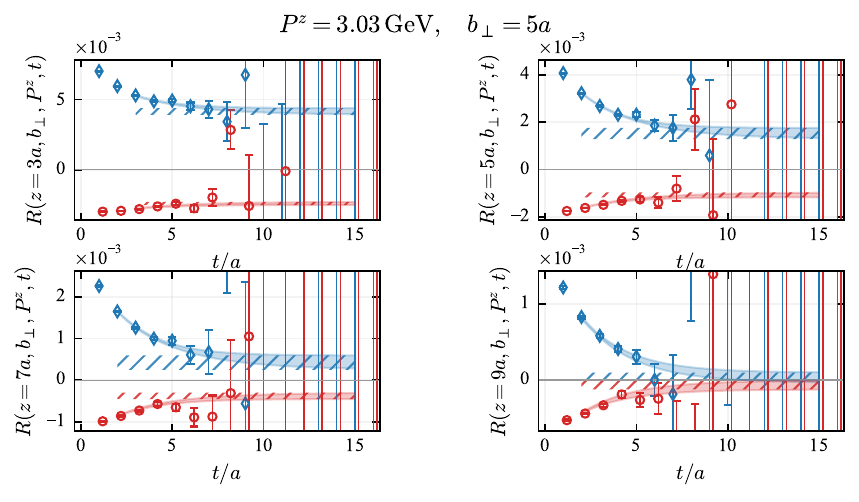}

    \caption{Extraction of the bare quasi-TMDWF matrix elements on ensemble G36P29 with $P^z=3.03$~GeV, $b_\perp=5a$, and $z=3,5,7$ and $9a$. Blue and red points denote the real and imaginary parts, and the dashed lines denote the plateau bands.}
    \label{fig:G36-bare-ME-pz6-b5-z3-9}
\end{figure*}

\begin{figure*}[htbp]
    \centering
    \includegraphics[width=\textwidth,
    trim= 0 0.2cm 0 0,
    clip]{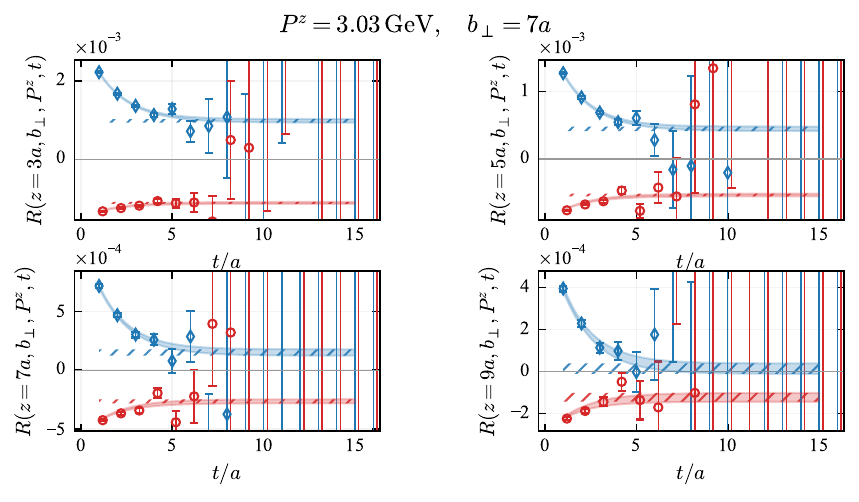}

    \caption{Extraction of the bare quasi-TMDWF matrix elements on ensemble G36P29 with $P^z=3.03$~GeV, $b_\perp=7a$, and $z=3,5,7$ and $9a$. Blue and red points denote the real and imaginary parts, and the dashed lines denote the plateau bands.}
    \label{fig:G36-bare-ME-pz6-b7-z3-9}
\end{figure*}

Similarly, we extract the bare quasi-TMDWF from the ratio of the nonlocal to the local two-point correlation functions,
\begin{align}
R(z,b_\perp,P^z,t)&=\frac{C_2^{\Gamma'}(t,z,b_\perp,P^z,\ell,\Gamma_\psi)}{C_2^{\Gamma'}(t,0,0,P^z,0,\Gamma_\psi)}\nonumber\\&=\widetilde{\Phi}^{(0)}(z,b_\perp,P^z)
\frac{1+d_0(z,b_\perp,P^z,\ell)e^{-\Delta Et}}{1+d_1e^{-\Delta Et}},
\end{align}
where $\widetilde{\Phi}^{(0)}(z,b_\perp,P^z)$ is the bare quasi-TMDWF in coordinate space, $\Gamma_\psi$ is the Dirac structure used for the quasi-TMDWF as in Ref.~\cite{Tan:2025ofx}, and $d_0$ and $d_1$ are free parameters that describe the excited-state contributions. As an example, the extraction of $\widetilde{\Phi}^{(0)}(z,b_\perp,P^z)$ on ensemble G36P29 for several momenta and transverse separations is shown in Figs.~\ref{fig:G36-bare-ME-pz4-b3-z3-9} to \ref{fig:G36-bare-ME-pz6-b7-z3-9}.

\begin{figure*}[htbp]
    \centering
    \includegraphics[width=\textwidth,height=\textheight,keepaspectratio]{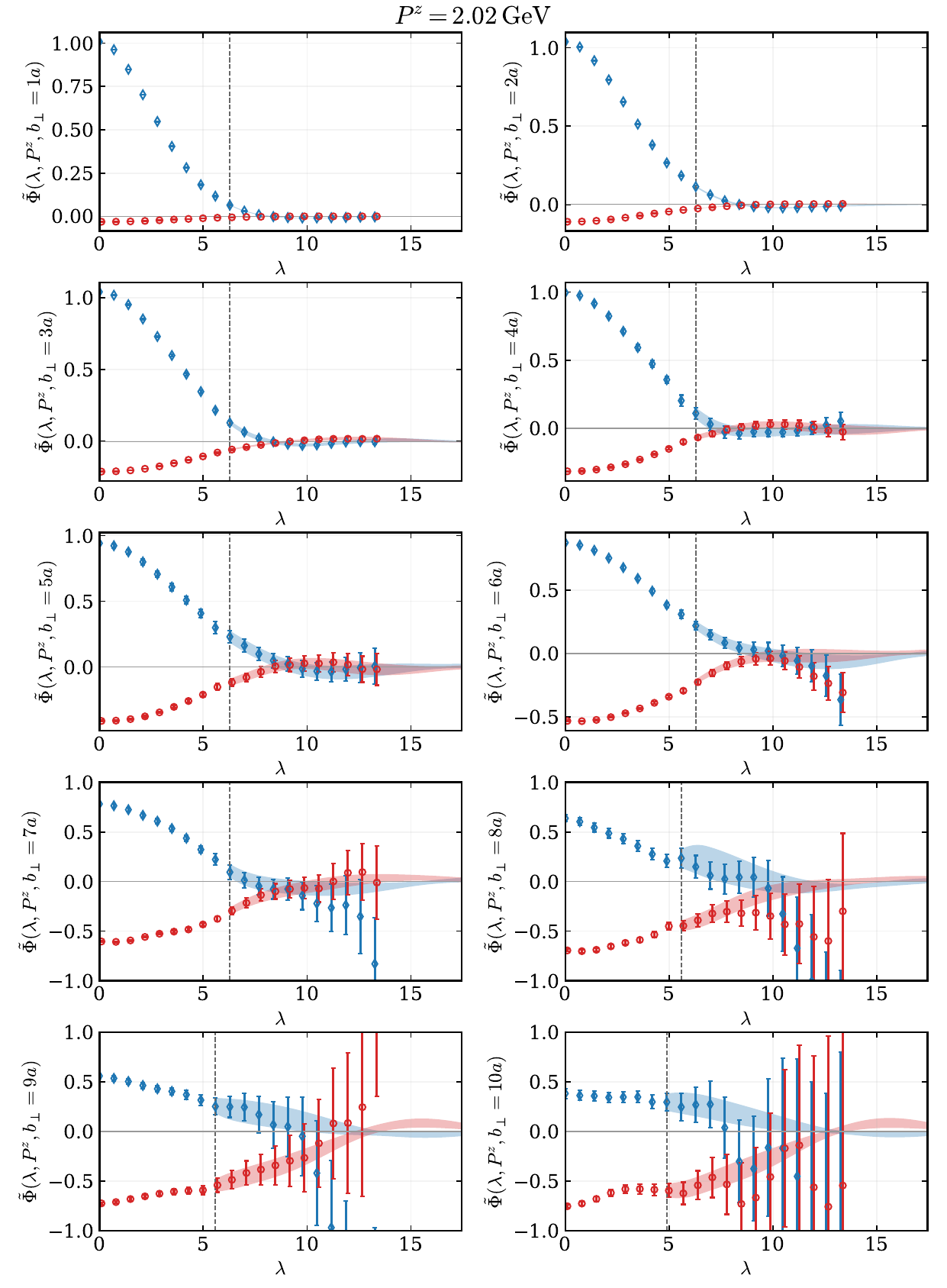}

    \caption{Extrapolation of the quasi-TMDWF in coordinate space as a function of $\lambda=zP^z$ on ensemble G36P29 with $P^z=2.02$~GeV. Blue and red points denote the real and imaginary parts.}
    \label{fig:G36-extra-pz4}
\end{figure*}

\begin{figure*}[htbp]
    \centering
    \includegraphics[width=\textwidth,height=\textheight,keepaspectratio]{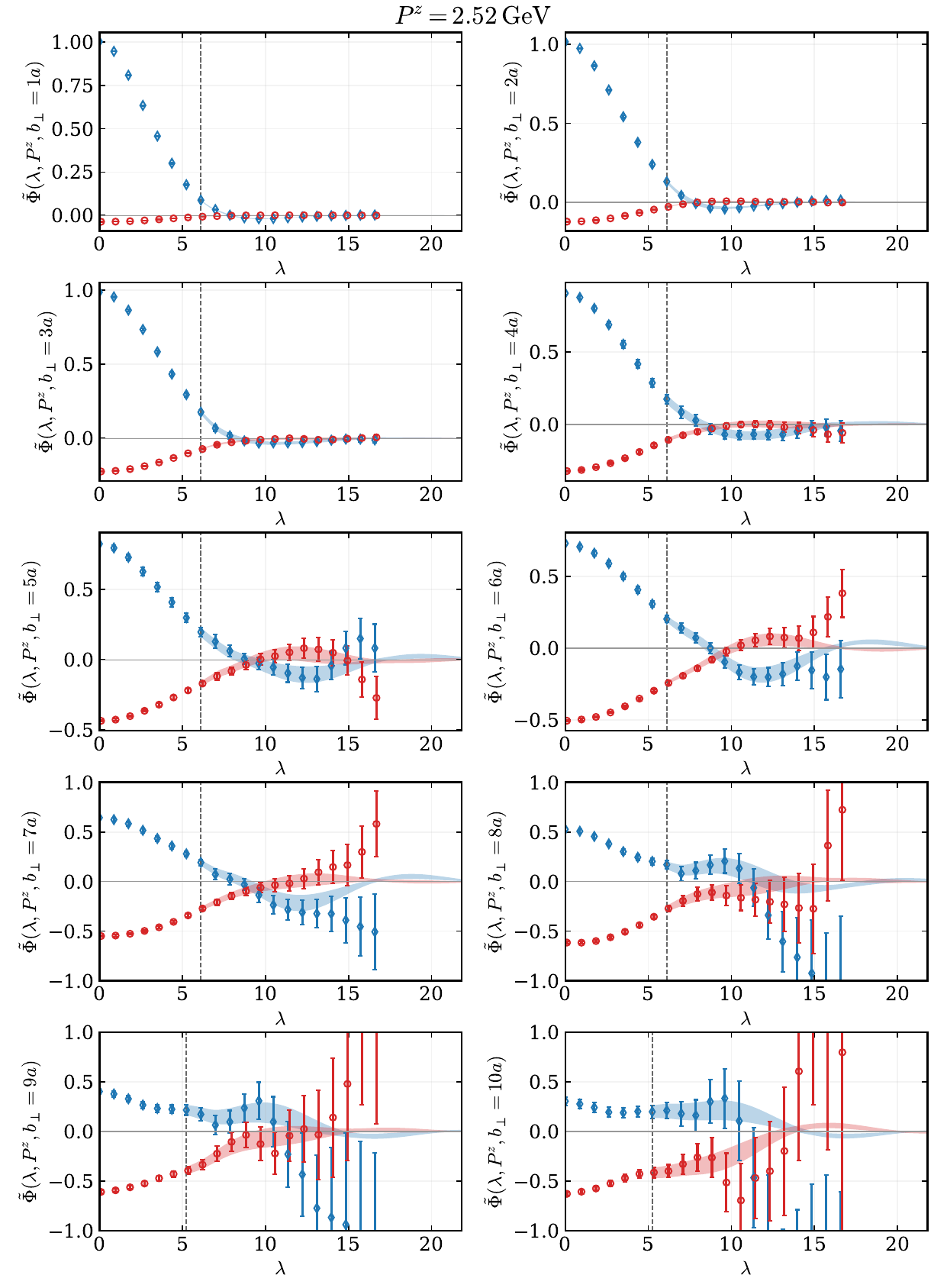}

    \caption{Extrapolation of the quasi-TMDWF in coordinate space as a function of $\lambda=zP^z$ on ensemble G36P29 with $P^z=2.52$~GeV. Blue and red points denote the real and imaginary parts.}
    \label{fig:G36-extra-pz5}
\end{figure*}

\begin{figure*}[htbp]
    \centering
    \includegraphics[width=\textwidth,height=\textheight,keepaspectratio]{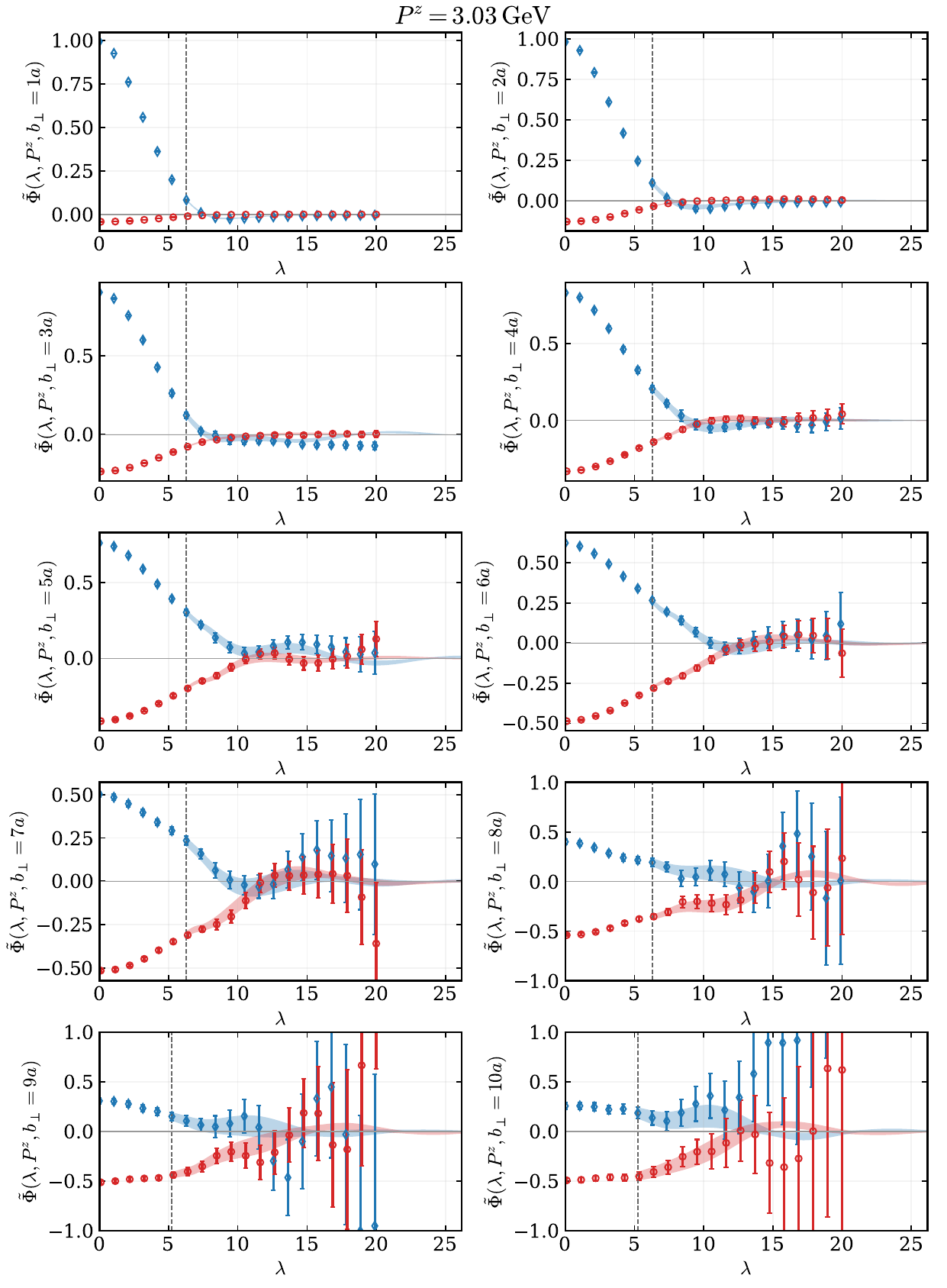}

    \caption{Extrapolation of the quasi-TMDWF in coordinate space as a function of $\lambda=zP^z$ on ensemble G36P29 with $P^z=3.03$~GeV. Blue and red points denote the real and imaginary parts.}
    \label{fig:G36-extra-pz6}
\end{figure*}

\clearpage
After renormalization, the coordinate-space matrix elements are extrapolated to large $\lambda=zP^z$ before the Fourier transform. This removes the unphysical oscillations caused by the finite range of the quasi-TMDWF, as shown in Figs.~\ref{fig:G36-extra-pz4} to \ref{fig:G36-extra-pz6}. Following Ref.~\cite{Tan:2025ofx}, the difference between the exponential-decay extrapolation~\cite{Ji:2020brr} and the form adopted for the pion distribution amplitude in Ref.~\cite{Ji:2026vir} is taken as a systematic uncertainty.

\begin{figure*}[htbp]
\centering
\includegraphics[width=\linewidth,height=0.9\textheight,keepaspectratio]{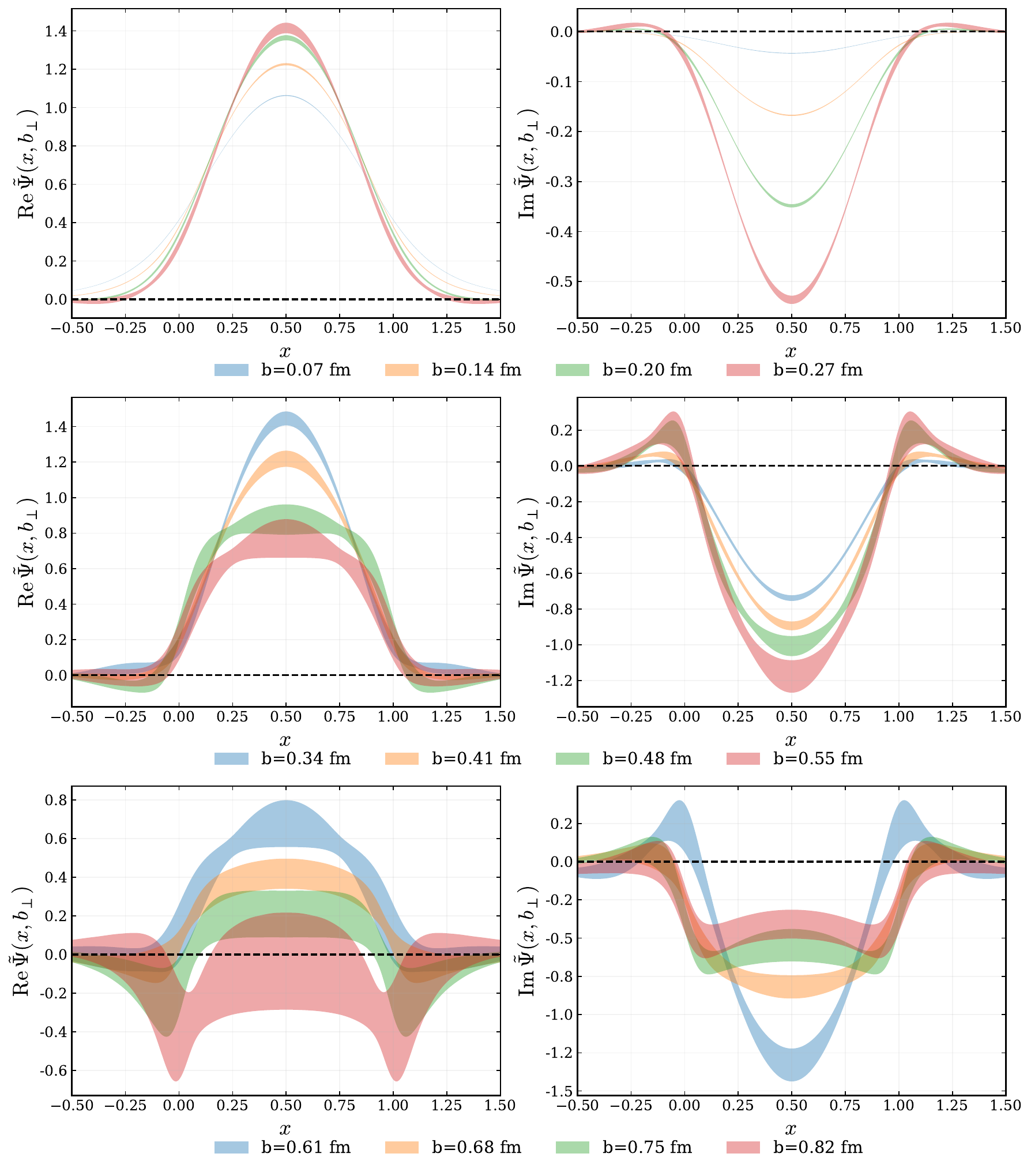}
\caption{Renormalized quasi-TMDWF on ensemble G36P29 with $P^z=2.02$~GeV and $b_\perp=[1,12]a$. The quasi-TMDWFs on the other ensembles are computed with the same setup as in Ref.~\cite{Tan:2025ofx}.}
\label{fig:g36_wf_p4_b1-12}
\end{figure*}

\begin{figure*}[htbp]
\centering
\includegraphics[width=\linewidth,height=0.9\textheight,keepaspectratio]{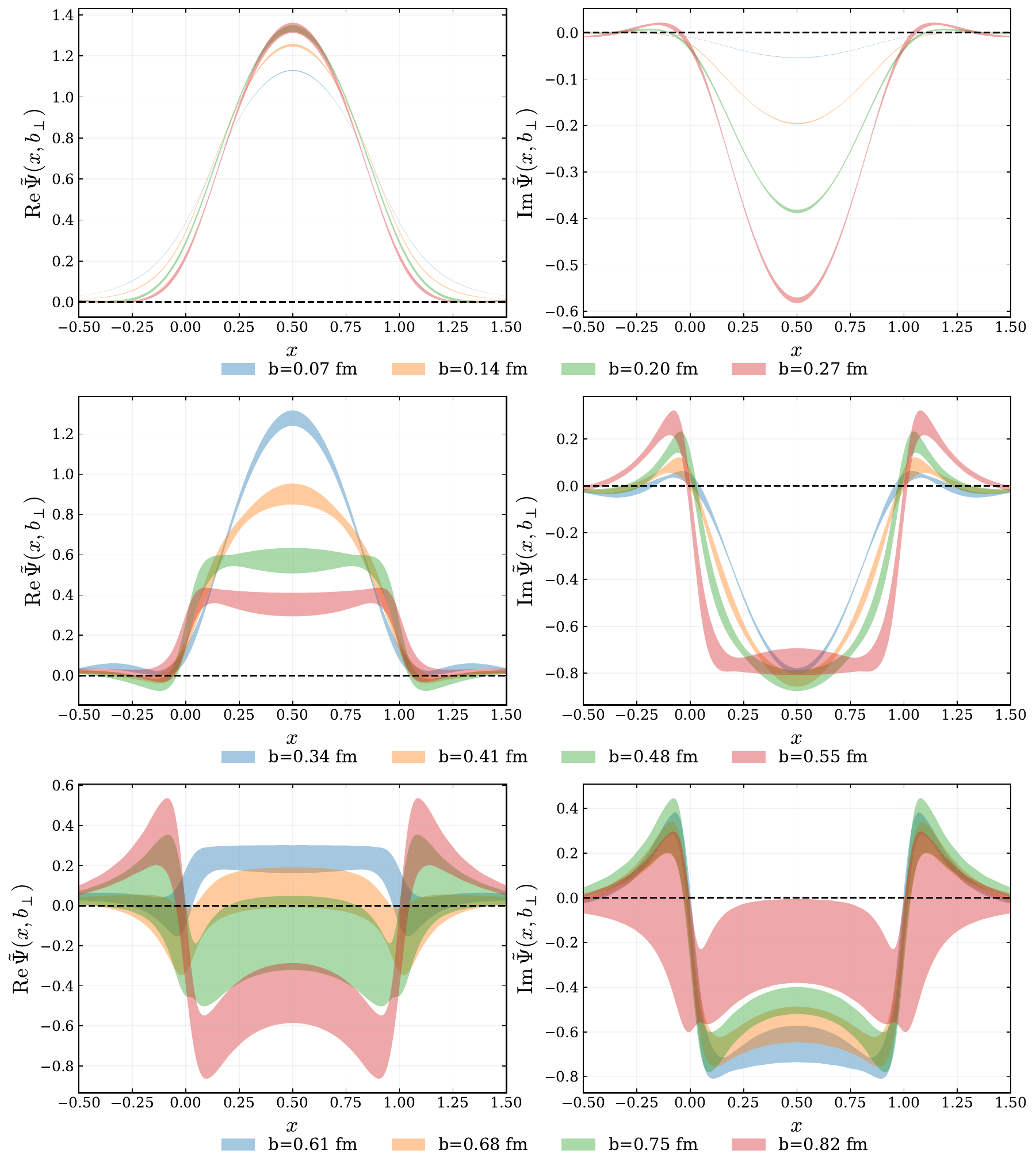}
\caption{Renormalized quasi-TMDWF on ensemble G36P29 with $P^z=2.52$~GeV and $b_\perp=[1,12]a$.}
\label{fig:g36_wf_p5_b1-12}
\end{figure*}

\begin{figure*}[htbp]
\centering
\includegraphics[width=\linewidth,height=0.9\textheight,keepaspectratio]{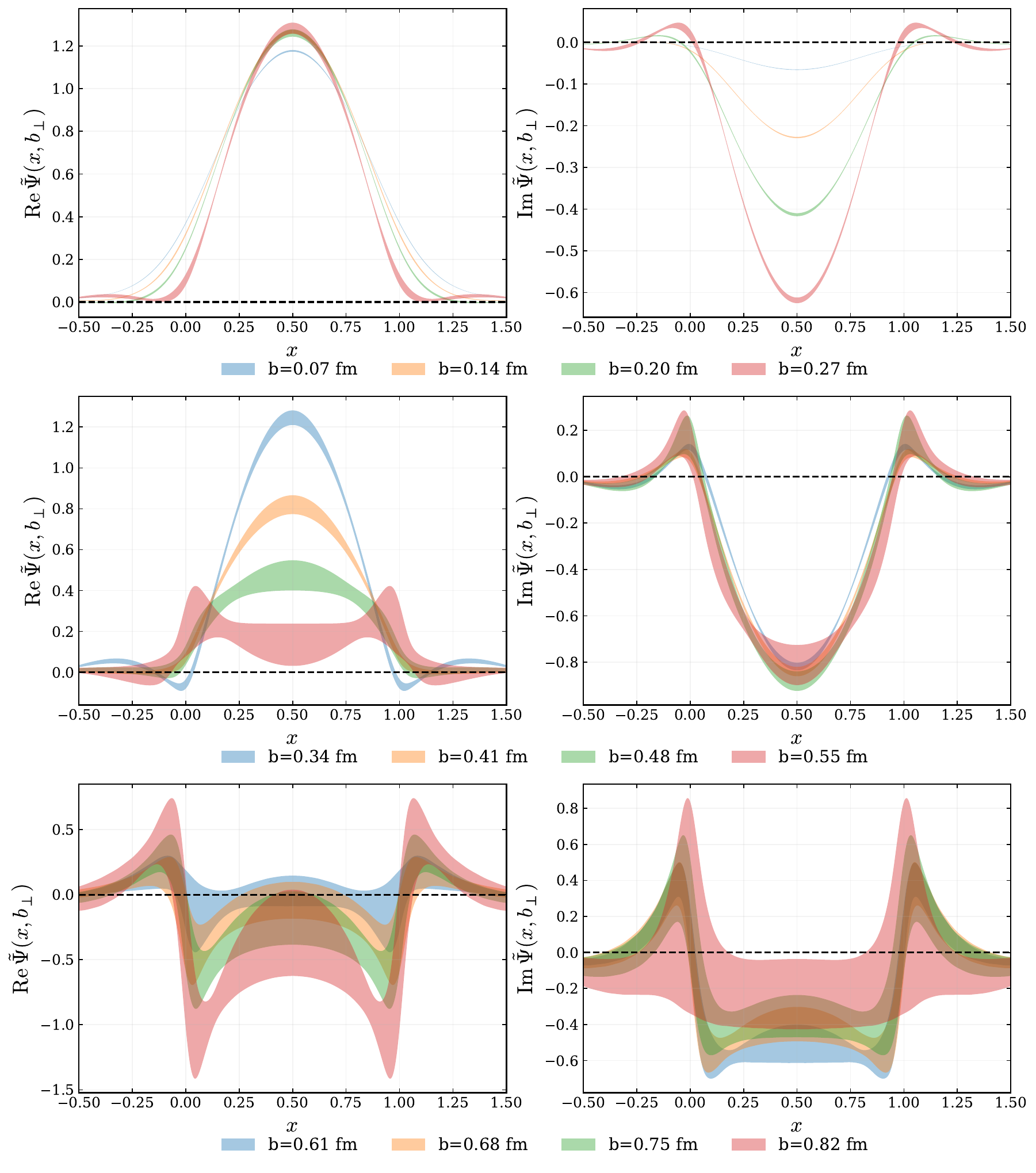}
\caption{Renormalized quasi-TMDWF on ensemble G36P29 with $P^z=3.03$~GeV and $b_\perp=[1,12]a$.}
\label{fig:g36_wf_p6_b1-12}
\end{figure*}

On the other five ensembles, the quasi-TMDWFs are computed with the same setup as in Ref.~\cite{Tan:2025ofx}. For the new ensemble G36P29, which provides an additional intermediate lattice spacing, we renormalize the bare quasi-TMDWF and Fourier transform it to momentum space following the same procedure. The quasi-TMDWFs on G36P29 are computed up to $b_\perp=15a$, covering the same transverse range as the form factor, and representative results for $b_\perp\le12a$ are shown in Figs.~\ref{fig:g36_wf_p4_b1-12} to \ref{fig:g36_wf_p6_b1-12}.

\subsection{\label{sec:soft}C. Extraction of the Intrinsic Soft Function}

\begin{figure*}[htbp]
\centering
\includegraphics[width=\linewidth]{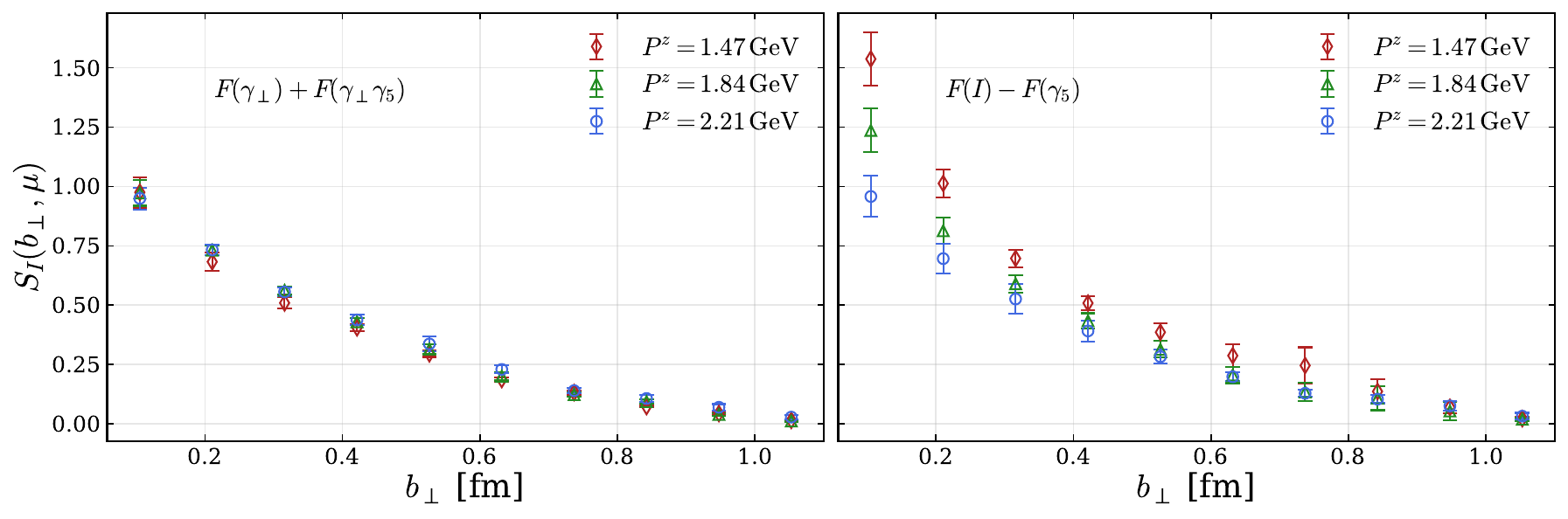}
\caption{Intrinsic soft function on ensemble C32P29 extracted from different current combinations. The left panel uses $F(\Gamma=\gamma_i)+F(\Gamma=\gamma_i\gamma_5)$ with $i=x,y$, and the right panel uses $F(\Gamma=I)-F(\Gamma=\gamma_5)$.} 
\label{fig:current_comparison}
\end{figure*}

We extract the intrinsic soft function from the renormalized form factor $F(b_\perp,P^z,\mu)$. In principle, the scalar ($\Gamma=I$) and pseudoscalar ($\Gamma=\gamma_5$) currents can also be used. However, they require additional ultraviolet renormalization and have nonzero anomalous dimensions, which lead to a large residual $P^z$ dependence at finite perturbative order~\cite{Deng:2022gzi,LatticePartonLPC:2023pdv}, as shown in Fig.~\ref{fig:current_comparison}. We therefore use the vector ($\Gamma=\gamma_i$) and axial-vector ($\Gamma=\gamma_i\gamma_5$) currents with $i=x,y$. The renormalized form factor $F$ is the sum of the bare form factors $F^{(0)}_{\gamma_i}$ and $F^{(0)}_{\gamma_i\gamma_5}$ over $i=x,y$, where the vector channel is multiplied by $Z_V^2/Z_A^2$, and $Z_V$ and $Z_A$ are the renormalization factors of the vector and axial-vector currents~\cite{CLQCD:2024yyn}. For the quasi-TMDWFs, we use the hybrid renormalization scheme of Ref.~\cite{Tan:2025ofx}, in which the Wilson-line power divergences are subtracted nonperturbatively, the remaining logarithmic divergence is treated by self-renormalization, and the result is converted to the $\overline{\rm MS}$ scheme.
At leading power, the form factor factorizes into the intrinsic soft function, the quasi-TMDWFs, and a perturbative matching kernel. On each ensemble, the intrinsic soft function is then obtained as
\begin{align}
    S_I^{\rm lat}&(b_\perp,\mu;a,m_\pi,P^z) = F(b_\perp,P^z,\mu) \nonumber \\
    &\times \biggl[ \int dx_1 dx_2 H^{+}(x_1,x_2,P^z,\mu) \nonumber \\
    &\times \widetilde{\Psi}^{+*}(x_1,b_\perp,-P^z)\widetilde{\Psi}^{+}(x_2,b_\perp,P^z) \biggr]^{-1} \!\! ,
    \label{eq:sm_SI_extraction}
\end{align}
that is, from the ratio of the form factor to the convolution in the square brackets. The superscript $+$ denotes the counterclockwise orientation of the staple-shaped Wilson line. The two orientations, denoted by $\pm$, give equivalent results for the intrinsic soft function~\cite{Deng:2022gzi,LatticeParton:2023xdl,LatticePartonLPC:2023pdv}, and we use only the $+$ orientation in this work.

The perturbative matching kernel has been calculated at one-loop order~\cite{Deng:2022gzi}. For both $\Gamma=\gamma_i$ and $\Gamma=\gamma_i\gamma_5$, it reads
\begin{align}
H^{\pm}(x_1, x_2,P^z,&\mu) = -\frac{1}{4N_c} \Biggl\{ 1 + \frac{\alpha_s C_F}{2\pi} \biggl[ \pi^2 - 4 \nonumber\\
&+\frac{1}{2} \ln^2 \left( -\frac{x_2}{x_1} \mp i0 \right) + \frac{1}{2} \ln^2 \left( -\frac{\bar{x}_2}{\bar{x}_1} \mp i0 \right) \nonumber\\
&+ \frac{1}{2} \ln \frac{16 x_1 x_2 \bar{x}_1 \bar{x}_2 (P^{z})^4}{\mu^4} \biggr] \Biggr\},
\label{eq:sm_kernel}
\end{align}
where $\bar{x}_1=1-x_1$ and $\bar{x}_2=1-x_2$, and the upper signs correspond to the $+$ orientation used in this work.

\begin{figure*}[htbp]
    \centering
    \includegraphics[width=\textwidth,height=0.9\textheight,keepaspectratio]{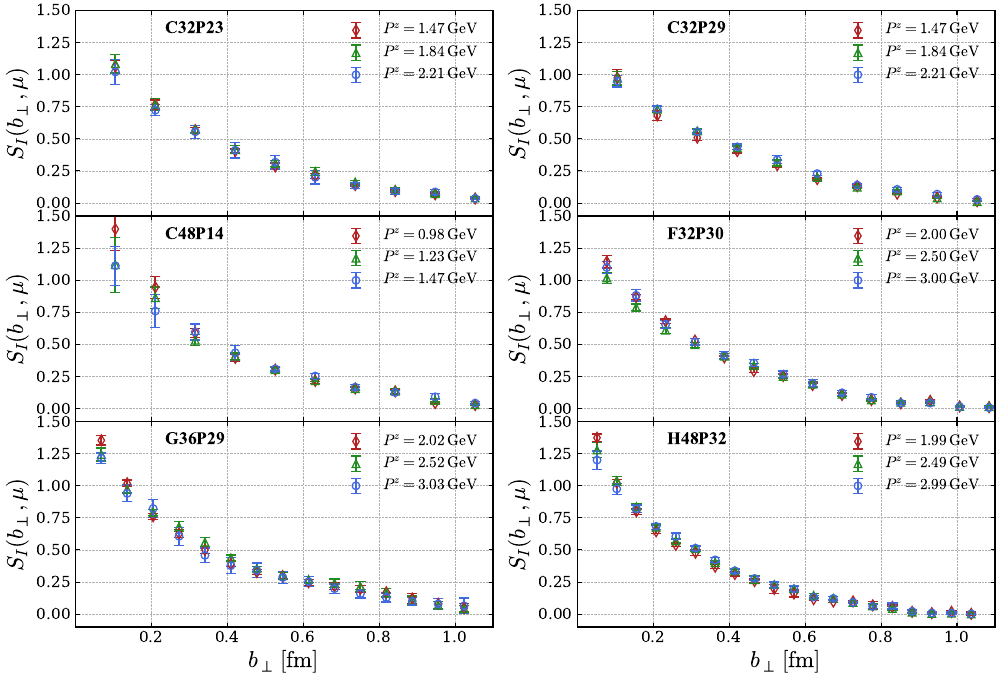}
    \caption{Intrinsic soft functions extracted on each ensemble. From left to right and top to bottom, the panels correspond to C32P23, C32P29, C48P14, F32P30, G36P29, and H48P32.}
    \label{fig:intrinsic-soft-function}
\end{figure*}

Figure~\ref{fig:intrinsic-soft-function} shows $S_I^{\rm lat}$ extracted on all ensembles, for three momenta and all transverse separations. Compared with the clear $P^z$ dependence of the form factors in Fig.~\ref{fig:form-factor-each-ensemble}, the intrinsic soft functions depend much more weakly on $P^z$. This shows that most of the $P^z$ dependence of the form factor is canceled by the quasi-TMDWFs and the matching kernel, which provides a check of the LaMET factorization. The remaining $P^z$ dependence is removed by the extrapolation to $P^z\to\infty$ described in Sec.~D.

\clearpage

\subsection{\label{sec:extrap}D. Details of the Infinite-Momentum, Continuum, and Physical-Mass Extrapolations}

\begin{figure}[htbp]
\centering
\includegraphics[ width=\linewidth
]{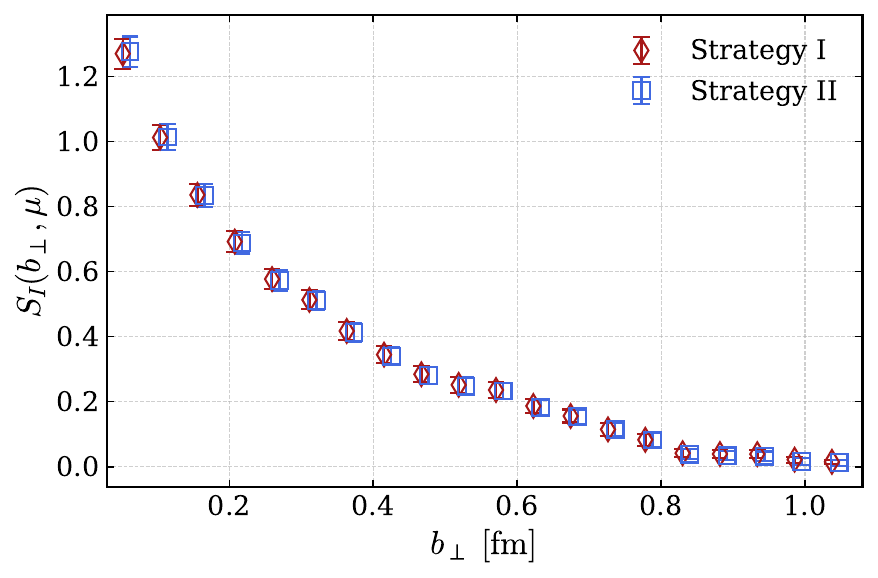}
\caption{Comparison of the intrinsic soft functions obtained from Strategy~I [$C(b_\perp)a^2$] and Strategy~II [$C_1(b_\perp)a+C_2(b_\perp)a^2$]. The two results agree within uncertainties.}
\label{fig:Strategy_comparison}
\end{figure}

\begin{figure*}[t]
\centering
\includegraphics[ width=\textwidth]{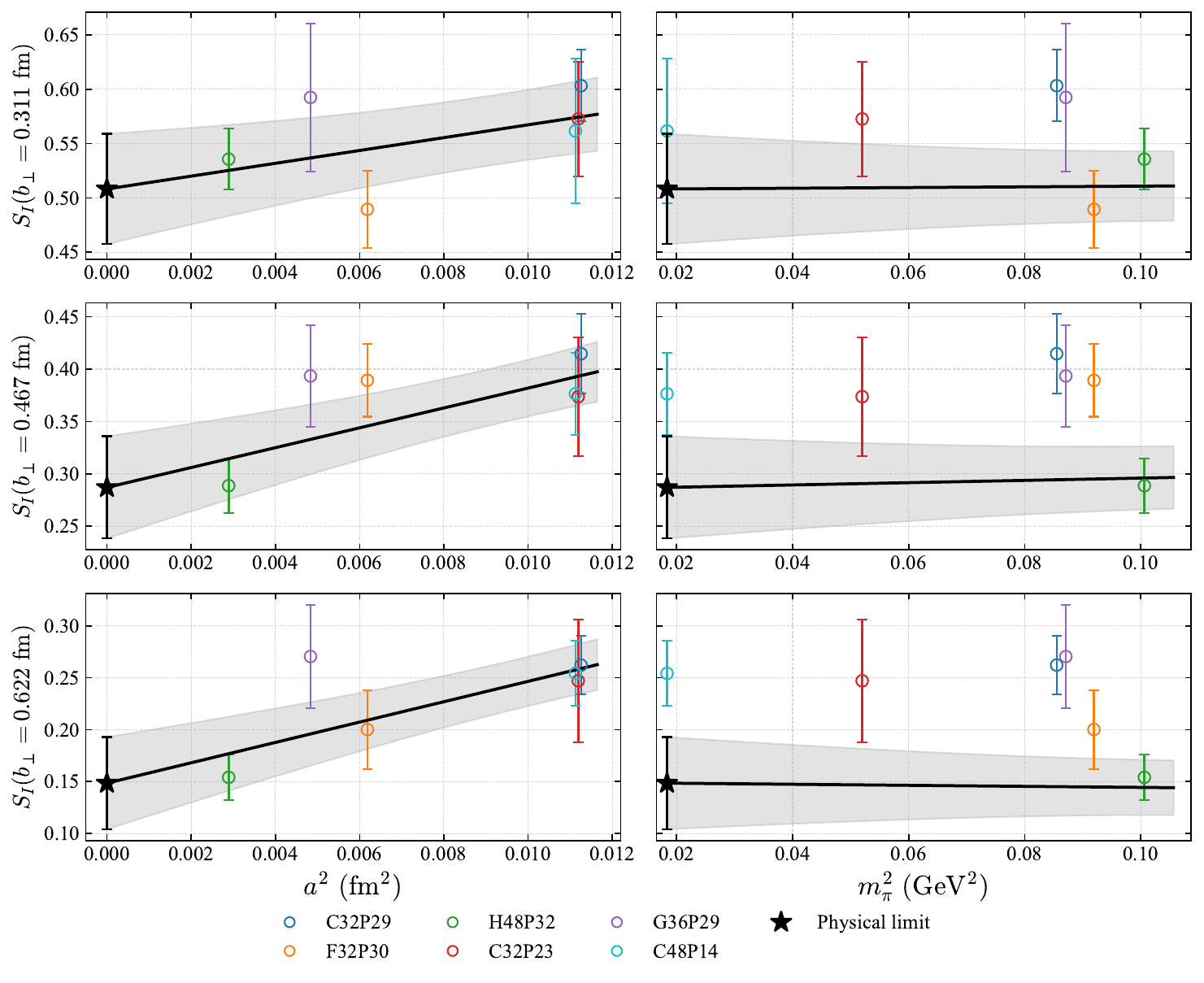}
\caption{Dependence of the intrinsic soft function on the lattice spacing (left) and pion mass (right). To remove finite-momentum effects, we first extrapolate the results of each ensemble separately to the infinite-momentum limit. The colored open markers denote the extrapolated results for individual ensembles, while the black star represents the result of a joint fit in lattice spacing and pion mass, extrapolated to the continuum limit and the physical point.
}
\label{fig:extrap_quality}
\end{figure*}

In the main text, we obtain the intrinsic soft function from a correlated joint fit at each $b_\perp$ to all ensembles and momenta, which extrapolates the results to the continuum limit, the physical point, and infinite momentum. We write the general fit form as
\begin{align}
S_I^{\rm lat}(b_\perp,\mu;a,m_\pi,P^z)=S_I(b_\perp,\mu)+\frac{A(b_\perp)}{(P^z)^2}&\nonumber\\
+B(b_\perp)\left[m_\pi^2-(m_\pi^{\rm phy})^2\right]+\Delta_{\rm disc}(a)&,
\label{eq:sm_general_fit}
\end{align}
where $A(b_\perp)$ describes the residual momentum dependence and $B(b_\perp)$ the residual pion-mass dependence. Since the pion mass is varied through the light-quark mass of the ensembles, the $B$ term includes both the external-state-mass effects in the power corrections and the sea-quark-mass dependence of $S_I$ itself. The term $\Delta_{\rm disc}(a)$ describes the discretization effects. Our baseline extrapolation, Strategy~I, uses $\Delta_{\rm disc}(a)=C(b_\perp)a^2$, which is the fit form used in the main text. Since the lattice actions used in this work suppress the $\mathcal{O}(a)$ discretization effects at tree level, we expect the cutoff dependence to be dominated by $\mathcal{O}(a^2)$ effects~\cite{CLQCD:2023sdb}. To estimate the systematic uncertainty from the choice of the continuum extrapolation form, we also use Strategy~II, which includes an additional linear term, $\Delta_{\rm disc}(a)=C_1(b_\perp)a+C_2(b_\perp)a^2$.

Since the lattice spacing gives the dominant residual dependence, we show in Fig.~\ref{fig:extrap_quality} the dependence on the lattice spacing and the pion mass at three representative transverse separations. For this illustration, the results on each ensemble are first extrapolated to $P^z\to\infty$ separately and then fitted in $a^2$ and $m_\pi^2$. The black lines and gray bands show the fitted $a^2$ dependence (left) and $m_\pi^2$ dependence (right), which connect the data on individual ensembles to the extrapolated results. The right panel shows a weak pion-mass dependence, consistent with the small $B$ term found in the main text. The extrapolated results are mainly determined by the data at the finest lattice spacings, where the discretization effects are small. This indicates that the discretization effects in our final result are under control.

Figure~\ref{fig:Strategy_comparison} compares the intrinsic soft functions obtained from Strategy~I and Strategy~II. The two results agree within uncertainties, and Strategy~II has larger uncertainties because it includes one more fit parameter. This agreement shows that the remaining $\mathcal{O}(a)$ effects are smaller than the statistical errors, which supports the use of Strategy~I in the main text. We take the difference between the extrapolated results of the two strategies as the discretization uncertainty, and include it in the total systematic uncertainty. The shifts associated with the $P^z\to\infty$ and physical-mass extrapolations are combined into the finite-momentum and mass uncertainty shown in Fig.~3 of the main text.

\end{document}